\documentclass[twocolumn,prd,nofootinbib,aps,floats,floatfix,amsmath,amssymb,secnumarabic]{revtex4} %

\usepackage{graphicx}
\usepackage[usenames,dvipsnames]{color}
\usepackage{amsmath,amssymb}
\usepackage{slashed}
\usepackage{verbatim}
\usepackage[colorlinks,citecolor=blue]{hyperref}
\def\sfrac#1#2{{\textstyle{#1\over #2}}}
\newcommand{\be}{\begin{equation}}
\newcommand{\ee}{\end{equation}}
\newcommand{\ba}{\begin{array}}
\newcommand{\ea}{\end{array}}
\newcommand{\bea}{\begin{eqnarray}}
\newcommand{\eea}{\end{eqnarray}}
\newcommand{\sss}{\scriptscriptstyle}
\newcommand{\R}{{\sss R}}

\newcommand{\nn}{\nonumber}
\renewcommand{\L}{{\sss L}}

\def\sfrac#1#2{{\textstyle{#1\over #2}}}

\def\beq{\begin{equation}}
\def\eeq{\end{equation}}
\def\roughly#1{\mathrel{\raise.3ex\hbox
{$#1$\kern-.75em\lower1ex\hbox{$\sim$}}}}

\def\bd{B^0}
\def\bs{B_s^0}
\def\bdbar{{\bar B}^0}
\def\bsbar{{\bar B}^0_s}

\def\btos{b \to s}
\def\sla#1{\raise.15ex\hbox{$/$}\kern-.57em #1}% Feynman slash

\def\bsmumu{b \to s \mu^+ \mu^-}
\def\bsnunubar{b \to s \nu {\bar\nu}}
\def\bsll{b \to s \ell^+ \ell^-}

\def\fbar{{\bar f}}
\def\qbar{{\bar q}}
\def \expt{{\rm expt}}
\def \s{\sqrt{2}}

\begin{document}
\leftline{CERN-TH-2017-025, UdeM-GPP-TH-17-254}
\title{\boldmath Hidden sector explanation of $B$-decay and
cosmic ray anomalies}
%\title{\boldmath $Z'$ explanation of $B$-decay anomalies, dark matter and AMS antiprotons}
\author{James M.\ Cline}
\email{jcline@physics.mcgill.ca}
\affiliation{CERN, Theoretical Physics Department, Geneva,
Switzerland}
\affiliation{Department of Physics, McGill University,
3600 Rue University, Montr\'eal, Qu\'ebec, Canada H3A 2T8}
\author{Jonathan M.\ Cornell}
\email{cornellj@physics.mcgill.ca}
\affiliation{Department of Physics, McGill University,
3600 Rue University, Montr\'eal, Qu\'ebec, Canada H3A 2T8}
\author{David London}
\email{london@lps.umontreal.ca}
\author{Ryoutaro Watanabe}
\email{watanabe@lps.umontreal.ca}
\affiliation{Physique des Particules, Universit\'e de Montr\'eal,
C.P. 6128, succ.\ centre-ville, Montr\'eal, QC, Canada H3C 3J7}

\begin{abstract} 

There are presently several discrepancies in $\bsll$ decays of $B$
mesons suggesting new physics coupling to $b$ quarks and leptons.  We
show that a $Z'$, with couplings to quarks and muons that can 
explain the $B$-decay anomalies, can also
couple to dark matter in a way that is consistent with its relic
abundance, direct detection limits, and hints of indirect detection.
The latter include possible excess events in
antiproton spectra recently observed by the AMS-02 experiment.
We present two models, having a heavy
(light) $Z'$ with $m_{Z'}\sim 600\,(12)\,$GeV and fermionic dark
matter with mass $m_\chi \sim 50\,(2000)\,$GeV, producing excess
antiprotons with energies of $\sim 10\, (300)$\,GeV.  The first model
is also compatible with fits for the galactic center GeV gamma-ray excess.

\end{abstract}
\maketitle

\section{Introduction}

At present, there are several measurements of $\bsll$ decays that
suggest the presence of physics beyond the standard model (SM):
\begin{itemize}

\item The LHCb Collaboration has measured the ratio $R_K \equiv {\cal
  B}(B^+ \to K^+ \mu^+ \mu^-)/{\cal B}(B^+ \to K^+ e^+ e^-)$, finding
  $R_K^\expt = 0.745^{+0.090}_{-0.074}~{\rm (stat)} \pm 0.036~{\rm
  (syst)}$ \cite{RKexpt}. Thus, a signal of lepton flavor
  nonuniversality at the level of 25\% was found, a deviation of
  $2.6\,\sigma$ from the SM prediction.

\item An angular analysis of $B \to K^* \mu^+\mu^-$ was performed by
  the LHCb \cite{BK*mumuLHCb1,BK*mumuLHCb2} and Belle
  \cite{BK*mumuBelle} Collaborations, and a discrepancy with the SM in
  the observable $P'_5$ \cite{P'5} was found. There are theoretical
  hadronic uncertainties in the SM prediction, but the deviation can
  be as large as $\sim 4\,\sigma$ \cite{BK*mumulatestfit1}.

\item The LHCb Collaboration has measured the branching fraction and
  performed an angular analysis of $B_s^0 \to \phi \mu^+ \mu^-$
  \cite{BsphimumuLHCb1,BsphimumuLHCb2}, finding a $3.5\,\sigma$
  disagreement with the predictions of the SM, which are based on
  lattice QCD \cite{latticeQCD1,latticeQCD2} and QCD sum rules
  \cite{QCDsumrules}.

\end{itemize}
What is particularly intriguing is that all these (independent)
discrepancies can be explained if there is new physics (NP) in
$\bsmumu$.  Numerous models have been proposed that generate the
correct NP contribution to $\bsmumu$ at tree level. They can
be put into two categories: those with a $Z'$ vector boson, and those
containing leptoquarks \cite{Hiller:2014yaa}.

Another indication of NP is dark matter (DM); the SM contains no
acceptable DM candidate.  Moreover the paradigm of WIMP (Weakly
Interacting Massive Particle) dark matter, which naturally obtains the
observed relic density through thermal processes, suggests that the DM
mass should be of the order of the electroweak scale. In light of
this, it is tempting to ask whether the NP responsible for the
$B$-meson anomalies may be connected to DM.  In particular, the new
particle that contributes to $\bsmumu$ could also be the mediator
connecting the DM to SM particles.  A simple possibility is that the
mediator is a $Z'$ associated with a $U(1)'$, under which the DM is
assumed to be charged.  We explore this idea here. Previous
work in this direction can be found in
Refs.~\cite{SSV,BDW,CFV,AGPQ}.   Our work has a different emphasis,
paying particular attention to recent hints of dark matter
annihilation contributing to the antiproton spectrum that has been
observed by the AMS-02 experiment \cite{AMS}.

Our starting point is the assumption that, at very high energies, the
flavor structure of the SM is gauged
\cite{gaugedflavor1,gaugedflavor2,gaugedflavor3,Guadagnoli:2011id}, and the SM group is
then extended by the maximal flavor group. It is further assumed that
this flavor group is spontaneously broken such that the only symmetry
left at the scale of $O$(TeV) is $U(1)'$. Only the left-handed
third-generation quarks and second-generation leptons in the flavor
basis are charged under this group. (Ref.~\cite{CFGI} has a similar
starting point, but assumes that the unbroken subgroups are $U(1)_q$
in the quark sector, and $U(1)_{\mu-\tau}$ in the lepton sector.)
The gauge boson associated with $U(1)'$ 
is denoted by 
$Z'$.  After electroweak symmetry breaking, when one transforms to the
mass basis, a flavor-changing coupling of the $Z'$ to $b_\L {\bar
  s_\L}$ is generated, leading to an effective $({\bar s_\L}
\gamma^\nu b_\L) \, ({\bar\mu_\L} \gamma_\nu \mu_\L)$ four-fermion
operator. This is used to explain the $\bsmumu$ anomalies.

In addition, we assume the presence of a DM fermion $\chi$ that is
charged under $U(1)'$. When $U(1)'$ is broken, a remnant global ${\cal
  Z}_2$ symmetry remains \cite{Z2sym1,Z2sym2}, ensuring the stability
of $\chi$. The $Z'$ acts as a mediator, enabling the annihilation
processes $\chi {\bar \chi} \to Z' \to f {\bar f}$ where $f$ is a SM
particle, mainly $b_\L$, $t_\L$, $\mu_\L$, $\nu_{\mu}$ in our model.
For light mediators, the process $\chi {\bar \chi} \to Z'Z'$ can
be dominant.

There are two variants of this $U(1)'$ model. In the first, the $Z'$
is heavy, $m_{Z'} = O$(TeV), the DM $\chi$ is a Dirac fermion of mass
$m_\chi \sim$ 30-70 GeV, and the $Z'$ couples to the $\chi$
vectorially. We demonstrate that values of
the model parameters can be found such that the NP contribution to
$\bsmumu$ explains the $B$ anomalies, while remaining consistent with
the constraints from $\bs$-$\bsbar$ mixing, $\bsnunubar$, neutrino
trident production, and LHC $Z'$ searches, as well as the DM
constraints from relic abundance, and direct and indirect
detection. The model also provides a tentative antiproton excess at
the 10 GeV energy scale \cite{Cuoco:2016eej,Cui:2016ppb}, as seen in
data from AMS-02.  An interesting feature of this model is that the
invisible contribution to the $Z'$ width from $Z'\to\chi\bar\chi$
allows it to escape the stringent LHC limits from dilepton searches
($Z'\to\mu\bar\mu$), that would otherwise exclude it.

In addition to the broad antiproton excess found at low (20-100\,GeV)
energies, there is also tentative evidence for a bump-like feature
near the end of the observed AMS-02 $\bar p$ spectrum. It has been
postulated that this feature could be explained by the production and
subsequent acceleration of $\bar p$ in supernova remnants
\cite{AMSexplain}, but here we consider a dark matter interpretation.
Ref.\ \cite{Cholis:2017qlb} showed that the annihilation of multi-TeV
DM into highly-boosted light mediators, that subsequently decay to
quarks, can produce the relatively narrow $\bar p$ peak around 300
GeV.  We find that a second variant of our model, with $m_{Z'} \cong$
12 GeV and quasi-Dirac DM of mass $m_\chi \cong 1950$ GeV, can give a
good fit to this observation, while evading bounds on direct detection
due to inelastic couplings of $Z'$ to the DM.  This model has strong
potential for discovery in upcoming LHC searches.

We begin in section \ref{model} by defining the model as regards the
$Z'$ couplings to SM particles.  In section \ref{flavorsec} we derive the
space of allowed parameters consistent the various flavor constraints.
Section \ref{DMsect} augments the model by coupling DM to the $Z'$.
Here we analyze the heavy and light $Z'$ variants of the model in some
detail, and demonstrate that it is possible to simultaneously explain the
$B$-decay anomalies and the antiproton excesses.  Conclusions are
given in sect.\ \ref{conclusions}.

\section{Model}
\label{model}

We start by defining the particle-physics model, at first ignoring its
couplings to dark matter, in order to address the anomalies in
$\bsmumu$.  We will later supplement the model (section \ref{DMsect})
with couplings to DM.

\subsection{Gauged flavor symmetries}

Refs.~\cite{gaugedflavor1,gaugedflavor2,gaugedflavor3,Guadagnoli:2011id} study the
effect of gauging the SM (quark or lepton) flavor symmetries. The
focus is principally to examine the relation between flavor-violating
effects and the Yukawa couplings, especially as regards avoiding
too-large flavor-changing neutral currents. An alternative to minimal
flavor violation \cite{MFV1,MFV2} is found. A crucial ingredient of
the analysis is the addition of new (chiral) fermions to cancel
anomalies.

In our model we assume that, at very high energies, the SM gauge group
$SU(3)_c \times SU(2)_L \times U(1)_Y$ is extended by the maximal
gauged flavor group $SU(3)_Q \times SU(3)_U \times SU(3)_D \times
SU(3)_\ell \times SU(3)_E \times O(3)_{\nu_R}$. Here $Q$ ($\ell$)
corresponds to the left-handed (LH) quarks (leptons), while $U$, $D$
and $E$ represent the right-handed (RH) up quarks, down quarks and
charged leptons, respectively. Three RH neutrinos are included in
order to generate neutrino masses via the seesaw mechanism, but are
otherwise unimportant for the model. We further assume that the flavor
group is spontaneously broken such that the only symmetry left at the
TeV scale is $U(1)'$. Only the LH third-generation quarks and
second-generation leptons are charged under this group.\footnote{As
  the underlying flavor group has been made anomaly-free by the
  addition of new fermions, this also resolves all anomaly problems
  associated with the $U(1)'$. Heavy fermions are required for the
  anomaly cancellation; we take these to have masses above the scales
  (TeV) in which we are interested. As a consequence, the only
  nonstandard fermion that couples to $Z'$ at lower energies is the
  dark matter.}  That is, $SU(3)_U \times SU(3)_D \times SU(3)_E
\times O(3)_{\nu_R}$ is broken completely, and $SU(3)_Q \times
SU(3)_\ell \to U(1)'$, with associated gauge boson $Z'$.

\subsection{Yukawa couplings}

At the TeV scale the Lagrangian is effective, and contains all
the terms left from integrating out the heavy fields. Consider the
Yukawa terms for the quarks, which connect LH and RH fields.
Since only LH third-generation quarks ($q_{3L}$) are charged under
$U(1)'$, any Yukawa term that does not involve $q_{3L}$ is as in the
SM: $\lambda_{ij} \qbar_{iL} H q_{jR} + h.c.$ ($i = 1,2, j = 1,2,3$).

On the other hand, Yukawa terms that involve $q_{3L}$ are of dimension
5: $[\lambda_j \qbar_{3L} H q_{jR} \Phi_q]/M +h.c.$ ($j = 1,2,3$), where
$M$ is the scale of some integrated-out particles, and $\Phi$ is a
scalar whose vacuum expectation value breaks $U(1)'$. (For the lepton
fields, the Yukawa terms are constructed similarly, except here the LH
second-generation leptons are treated like the LH third-generation
quarks.) Thus, when $\Phi$ gets a VEV, the Lagrangian contains the SM
terms, along with the $Z'$ couplings to SM particles, plus higher
dimension  nonrenormalizable terms that can be neglected.  At this
scale the SM terms include all the Yukawa couplings, $\lambda_{ij}
\fbar_{iL} H f_{jR} + h.c.$ ($i,j = 1,2,3$).

The simplest UV completion requires the introduction of heavy
isosinglet vectorlike quarks $T,B$, lepton $L$ and scalars
$\Phi_q,\Phi_l$ with U(1)$'$ 
charges $g_q$ and $g_l$ respectively, that match those of the SM
doublets $Q_{3,\L}$ and
$L_{2,\L}$.  Then the renormalizable terms
\bea
	{\cal L} &=& y'_{b} \bar Q_{3,\L} H B_\R + y'_{t} \bar Q_{3,\L} \tilde H 
	T_\R + y'_\mu \bar L_{2,\L} H L_\R\nonumber\\
	&+& \eta_{b,i} \bar B_\L \Phi_q d_{i,\R} + 
	\eta_{t,i} \bar T_\L \Phi_q u_{i,\R} +
	\eta_{\mu,i} \bar L_\L \Phi_l e_{i,\R}\nonumber\\
	&+& M_t \bar T T + M_b \bar B B + M_\mu \bar L L 
\label{eq1}
\eea
generate the dimension-5 Yukawa interactions after the heavy fermions
are integrated out.  The corresponding SM Yukawa couplings that are
most relevant for this study are
\bea
	\lambda_{tt} &=& y'_t \eta_{t,t} {\langle\Phi_q\rangle\over M_t}\nonumber\\
	\lambda_{bb} &=& y'_b \eta_{b,b} {\langle\Phi_q\rangle\over M_b}\nonumber\\
	\lambda_{bs} &=& y'_b \eta_{b,s} {\langle\Phi_q\rangle\over M_b}
\label{dim5Yuk}
\eea
Assuming that $\langle\Phi_q\rangle\sim M_t$, it is possible to 
generate a large enough top quark Yukawa coupling as long as
$y'_t\sim \eta_{t,t}\sim 1$.  The quark mixing needed to get 
the $b\to s$ transitions from $Z'$ exchange will be controlled by
$\eta_{b,s}/\eta_{b,b}$.

Since the current limit on vector-like isosinglet quarks is
$M > 870\,$GeV \cite{ATLAS:2017lvm}, the VEV $\langle \Phi_q\rangle$ contributes 
of order $(870\times g_q)$\,GeV  to the $Z'$ mass.  
We will find that satisfying flavor and dark matter constraints
requires $g_q\cong 0.4\, m_{Z'}/$TeV, which
is too small for this to be the sole 
contribution to $m_{Z'}$.  The rest must either come from $\langle
\Phi_l\rangle$, or from an additional dark scalar field that we will
introduce in a scenario with a light $Z'$.  Since the largest Yukawa
coupling in the lepton sector that must be generated by
$\langle\Phi_l\rangle$ is $\lambda_{\mu\mu}$, we have the freedom to
choose $\langle\Phi_l\rangle\ll \langle \Phi_q\rangle$, and we will
make this assumption in the light $Z'$ scenario to avoid too large
contributions to $m_{Z'}$.

\subsection{Four-fermion operators}

In the gauge basis, the Lagrangian describing the couplings of the
$Z'$ to fermions is
\bea
\label{Z'couplings}
\Delta {\cal L}_{Z'} & = & J^\mu Z'_\mu ~, \\
{\rm where} \qquad J^\mu & = & 
g_q ({\bar \psi}'_q \gamma^\mu P_L \psi'_q) + g_l ({\bar \psi}'_\ell \gamma^\mu P_L \psi'_\ell) ~. 
\eea
Here $\psi'_q$ ($\psi'_\ell$) represents both $t$ and $b$ ($\nu_\mu$
and $\mu^-$) fields, and the primes indicate the gauge basis. $g_q =
g_1 Q_q$ and $g_l = g_1 Q_\ell$ are the couplings of the $Z'$ to
quarks and leptons, respectively ($g_1$ is the $U(1)'$ coupling
constant, and $Q_q$ and $Q_\ell$ are the $U(1)'$ charges of quarks and
leptons). Once the heavy $Z'$ is integrated out, we obtain the
following effective Lagrangian containing 4-fermion operators:
\bea
{\cal L}_{Z'}^{eff} & = & -\frac{1}{2 m_{Z'}^2} J_\mu J^\mu \nn\\
& \supset & -\frac{g_q g_l}{m_{Z'}^2} ({\bar \psi}'_q \gamma_\mu P_L \psi'_q) 
({\bar \psi}'_\ell \gamma^\mu P_L \psi'_\ell) \nn\\
&& \hskip0.5truecm -~\frac{g_q^2}{2 m_{Z'}^2} ({\bar \psi}'_q \gamma_\mu P_L \psi'_q) 
({\bar \psi}'_q \gamma^\mu P_L \psi'_q) \nn\\
&& \hskip0.5truecm -~\frac{g_l^2}{2 m_{Z'}^2} ({\bar \psi}'_\ell \gamma_\mu P_L \psi'_\ell) 
({\bar \psi}'_\ell \gamma^\mu P_L \psi'_\ell) ~. 
\label{4fermionops}
\eea
The first 4-fermion operator (two quarks and two leptons) is relevant
for $\bsll$ and $\bsnunubar$ decays, the second operator (four quarks)
contributes to processes such as $\bs$-$\bsbar$ mixing, and the third
operator (four leptons) contributes to neutrino trident production and
$Z\to 4\mu$.

In order to obtain the operators involving the physical fields, we
must transform the fermions to the mass basis. We make the
{approximation
that the gauge and mass eigenstates are the same for all fermions
except the LH up- and down-type quarks. In the lepton sector, this
holds if neutrino masses are neglected.  For the quarks, it would be a
good approximation if $\lambda_{sb}$, which comes from the usual
dimension-4 Yukawa interaction, happens to be much smaller than
$\lambda_{bs}$ in eq.\ (\ref{dim5Yuk}). In this case the mixing angle
between 2nd and 3rd generation left-handed quarks is approximately
$\theta_\L \cong \eta_{b,s}/\eta_{b,b}$ while that of their
right-handed counterparts is smaller by a factor of $\sim m_s/m_b$.
In the following we therefore ignore $\theta_\R$.}

 In transforming from the
gauge basis to the mass basis, we then have
\beq
u'_L = U u_L ~,~~ d'_L = D d_L ~,
\label{transformations}
\eeq
where $U$ and $D$ are $3\times 3$ unitary matrices and the spinors
$u^{(\prime)}$ and $d^{(\prime)}$ include all three generations of
fermions. The CKM matrix is given by $V_{CKM}=U^\dagger D$.

For the $B$ anomalies, we are particularly interested in the decay
$\bsmumu$, i.e., the $Z'$ must couple to ${\bar s}b$ in the mass
basis. If the $Z'$ also couples to ${\bar d}s$ (${\bar d}b$), there
are stringent constraints from $K^0$-${\bar K}^0$ ($\bd$-$\bdbar$)
mixing. To avoid this, we assume that the $D$ transformation involves
only the second and third generations \cite{EffFT_3rdgen,RKRDmodels}:
\beq
D =
\left(
\begin{array}{ccc}
1 & 0 & 0 \\
0 & \cos\theta_D & \sin\theta_D \\
0 & -\sin\theta_D & \cos\theta_D
\end{array}
\right)
\label{Ddef}
\eeq
where $\theta_D = \theta_\L \cong \eta_{b,s}/\eta_{b,b}$ as mentioned
above.
With this transformation, for the down-type quarks, couplings
involving the second generation (possibly flavor-changing) are
generated in the mass basis. (For the up-type quarks, the first
generation can also be involved.)

Now, we are interested in $\btos$ transitions in the mass basis, and
these can arise through the exchange of a $Z'$. Applying the above
transformation to Eq.~(\ref{4fermionops}), we find the following. The
4-fermion operator applicable to $\bsmumu$ or $\bsnunubar$ is
\beq
\frac{g_q g_l}{m_{Z'}^2} \, \sin\theta_D \cos\theta_D \, ({\bar s} \gamma_\mu P_L b) ({\bar L} \gamma^\mu P_L L) ~.
\label{bsllNP}
\eeq
For $\bs$-$\bsbar$ mixing, the relevant operator is
\beq
- \frac{g_q^2}{2 m_{Z'}^2} \, \sin^2\theta_D \cos^2\theta_D \, ({\bar s} \gamma^\mu P_L b) ({\bar s} \gamma^\mu P_L b) ~.
\label{BsmixNP}
\eeq

\subsection{\boldmath $Z' d{\bar d}$ and $Z' u{\bar u}$ Couplings}

Although our immediate concern is $b\to s$ transitions, the small couplings
of $Z'$ to light quarks induced by mixing in our model will be
relevant later on, for the direct detection of dark matter.
 Because the $D$
transformation involves only the second and third generations
[Eq.~(\ref{Ddef})], the $Z' d{\bar d}$ coupling vanishes. Using
$V_{CKM}=U^\dagger D$, the $Z'$ coupling to LH up-type quarks is given
by
\beq
M = U^\dagger
\left(
\begin{array}{ccc}
0 & 0 & 0 \\
0 & 0 & 0 \\
0 & 0 & 1
\end{array}
\right)
U =
V_{CKM} D^\dagger 
\left(
\begin{array}{ccc}
0 & 0 & 0 \\
0 & 0 & 0 \\
0 & 0 & 1
\end{array}
\right)
D V_{CKM}^\dagger ~.
\eeq
The $Z' u{\bar u}$ coupling is then given by
\bea
M_{11} & = & |V_{us}|^2 \sin^2\theta_D - 2 \, {\rm Re}(V_{us} V_{ub}^*) \sin\theta_D \cos\theta_D \nn\\
&& \hskip1truecm +~|V_{ub}|^2 \cos^2\theta_D ~.
\label{Z'uu}
\eea
For very small $\theta_D$ such that $\sin\theta_D\cong\theta_D$ and
$\cos\theta_D\cong 1$, and neglecting the phase in $V_{us}
V_{ub}^*$, we can estimate 
$M_{11} \sim |V_{ub}-\theta_D V_{us}|^2$.

\section{Flavor constraints}
\label{flavorsec}

Here we determine the allowed values of $\theta_D$ versus $g_q
g_l/m_{Z'}^2$ that can explain the $\bsmumu$ anomalies, while
respecting constraints from $\bs$-$\bsbar$ mixing, $\bsnunubar$,
neutrino trident production, $Z\to 4\mu$ decays, and the muon
anomalous magnetic moment.

\subsection{\boldmath $\bsmumu$}

$\bsmumu$ transitions are described by the effective Hamiltonian
\bea
H_{\rm eff} &=& - \frac{\alpha G_F}{\s \pi} V_{tb} V_{ts}^*
      \sum_{a = 9,10} ( C_a O_a + C'_a O'_a ) ~, \nn\\
O_{9(10)} &=& [ {\bar s} \gamma_\mu P_L b ] [ {\bar\mu} \gamma^\mu (\gamma_5) \mu ] ~,
\label{Heff}
\eea
where the primed operators are obtained by replacing $L$ with $R$. The
Wilson coefficients $C^{(\prime)}_a$ include both SM and NP
contributions. In Ref.~\cite{BK*mumulatestfit1}, a global analysis of
the $\bsll$ anomalies was performed for both electron and muon decay
modes, including data on $B \to K^{(*)} \mu^+ \mu^-$, $B \to K^{(*)}
e^+ e^-$, $\bs \to \phi \mu^+ \mu^-$, $B \to X_s \mu^+ \mu^-$, $b \to
s \gamma$ and $\bs \to \mu^+ \mu^-$.  Theoretical hadronic
uncertainties were taken into account, and it was found that there is
a significant disagreement with the SM, possibly as large as
$4\sigma$. This discrepancy can be explained if there is NP in $b \to
s \mu^+ \mu^-$. There are four possible explanations, each having
roughly equal goodness-of-fits, but the one that interests us is
$C_9^{\rm NP} = - C_{10}^{\rm NP} < 0$.  According
to the fit, the allowed $3\sigma$ range for the Wilson coefficients is
\beq
-1.12 \le C_9^{\rm NP} = -C_{10}^{\rm NP} \le -0.18 ~.
\label{C9bounds}
\eeq

In our model, $\bsmumu$ transitions are given by the effective Hamiltonian
\begin{align}
 &
 H_{\rm eff}(\bsmumu) \notag \\
 & 
 = \left( - {\alpha G_F \over \sqrt 2 \pi} V_{tb} V_{ts}^* C_9^\text{SM} + \frac{g_q g_l}{2m_{Z'}^2} \, \sin\theta_D \cos\theta_D \right) \notag \\
 & 
 ~~~~\times \left(\bar s \gamma^\mu P_L b \right) \left( \bar\ell_i \gamma_\mu (1-\gamma^5) \ell_j \right)  ~,
\end{align}
where the SM contribution, $C_9^\text{SM} (= -C_{10}^\text{SM}) \simeq
0.94$ \cite{Bobeth:2013uxa}, encodes a loop suppression. This leads to
\bea
C_9^{\rm NP} 
&=& -C_{10}^{\rm NP} \nn\\
&=& {\pi \over \sqrt 2 \alpha G_F V_{tb}V_{ts}^*} \,
\frac{g_q g_l}{m_{Z'}^2} \, \sin\theta_D \cos\theta_D 
\label{C9NP}
\eea
in $\bsmumu$, while there is no NP contribution to $b \to s
e^+e^-$. Eq.~(\ref{C9bounds}) then constrains the combination of
theoretical parameters $\theta_D\,g_q g_l/m_{Z'}^2$ in the limit of
small $\theta_D$.

\subsection{\boldmath $\bs$-$\bsbar$ mixing}

In our model, $\bs$-$\bsbar$ mixing is described by the effective
Hamiltonian
\bea
 H_{\rm eff} &=& \left( N C_{VLL}^\text{SM} + \frac{g_q^2}{2 m_{Z'}^2} \, \sin^2\theta_D \cos^2\theta_D \right) \nn\\
&& ~~~~\times ({\bar s} \gamma^\mu P_L b)\,({\bar s} \gamma_\mu P_L b)~,
\eea
where $N = (G_F^2 m_W^2/16\pi^2) (V_{tb} V_{ts}^*)^2$ (the SM
contribution is produced via a box diagram), and $C_{VLL}^\text{SM}
\simeq 4.95$~\cite{RKRDmodels}. The mass difference in the $B_s$
system is then given by
\begin{align}
 \Delta M_s = 
 & \frac{2}{3} m_{B_s} f_{B_s}^2 \hat B_{B_s} \notag \\
 & \times \left | N C_{VLL}^{\rm SM} + \frac{g_q^2}{2 m_{Z'}^2} \, \sin^2\theta_D \cos^2\theta_D \right | ~.
\label{DeltaMstot}
\end{align}
The SM prediction is \cite{RKRDmodels}
\beq
\Delta M_s^{\rm SM} = (17.4 \pm 2.6)~{\rm ps}^{-1} ~.
\eeq
This is to be compared with the experimental measurement \cite{HFAG}
\beq
\Delta M_s = (17.757 \pm 0.021)~{\rm ps}^{-1} ~,
\eeq
leading to a constraint on $\theta_D^2\, g_q^2/m_{Z'}^2$ for
$\theta_D\ll 1$.

In the SM, the weak phase of $\bs$-$\bsbar$ mixing is predicted to be
very small: $\varphi_s = -0.03704 \pm 0.00064$
\cite{Charles:2004jd,Hocker:2001xe}. The present measurement of this
quantity is $\varphi_s^{c{\bar c}s} = -0.030 \pm 0.033$ \cite{HFAG}.
Although these values are consistent with one another, the
experimental error is large, allowing for a significant NP
contribution. This then raises the question: could the present $Z'$
model give a large contribution to $\varphi_s$? Unfortunately, the
answer is no. The $Z'$ contribution to $\bs$-$\bsbar$ mixing is given
in Eq.~(\ref{DeltaMstot}). It can include a weak phase only if $g_q$
is complex. However, from Eq.~(\ref{Z'couplings}), we see that, since
the coupling is self conjugate, the coupling constant $g_q$ is real.
Thus, if a future measurement of $\varphi_s^{c{\bar c}s}$ were to find
a sizeable deviation from the SM, it could not be accomodated in our
model.

\subsection{\boldmath $\bsnunubar$}

In our model, the effective Hamiltonian for $\bsnunubar$ is
\begin{align}
 &
 H_{\rm eff}(b \to s \nu_\mu\bar\nu_\mu) \notag \\
 &
 = 
 \left( - {\alpha G_F \over \sqrt 2 \pi} V_{tb} V_{ts}^* C_L^\text{SM} + \frac{g_q g_l}{2m_{Z'}^2} \, \sin\theta_D \cos\theta_D  \right)\, \notag \\
 &
 ~~~~\times\left(\bar s \gamma^\mu P_L b \right) \left( \bar\nu_\mu \gamma_\mu (1-\gamma^5) \nu_\mu \right)  \,, 
\end{align}
where the SM loop function is $C_L^\text{SM} \simeq -6.60$.  The NP
contribution can be constrained by the 90\% C.L. upper limits of
$\mathcal{B}(B^+ \to K^+ \nu\bar\nu) \le 1.7 \times 10^{-5}$,
$\mathcal{B}( B^+ \to K^{*+} \nu\bar\nu) \le 4.0 \times 10^{-5}$, and
$\mathcal{B}( B^0 \to K^{*0} \nu\bar\nu) \le 5.5 \times 10^{-5}$,
given by the BaBar and Belle
Collaborations~\cite{BKnunubarBaBar,BKnunubarBelle}.

Comparing the experimental upper limits with the SM predictions, the
resulting constraint (including theoretical uncertainties) is
\cite{Buras:2014fpa}
\begin{align}
 \frac{ 2 |C_L^{\rm SM}|^2 + |C_L^{\rm SM} + C_L^{\rm NP}|^2}{3|C_L^{\rm SM}|^2} \lesssim 5 \,, 
% 13\, \text{Re}[C_L^{\rm NP}] + |C_L^{\rm NP}|^2 \leq 473 \,, 
\end{align}
with 
\begin{align}
 C_L^{\rm NP} =  {\pi \over \sqrt 2 \alpha\, G_F V_{tb}V_{ts}^*} \,\frac{g_q g_l}{m_{Z'}^2} \, \sin\theta_D \cos\theta_D \,. 
\end{align}
 This has the same form as the NP contribution to $\bsmumu$
[Eq.~(\ref{C9NP})]. However, as we will see below, the constraint from
$\bsnunubar$ is quite a bit weaker than that from $\bsmumu$.

\subsection{Neutrino trident production}

A further constraint arises due to the effect of the $Z'$ boson on the
production of $\mu^+\mu^-$ pairs in neutrino-nucleus scattering,
$\nu_\mu N \to \nu_\mu N \mu^+ \mu^-$ (neutrino trident production).
At leading order, this process is effectively $\nu_\mu \gamma \to
\nu_\mu \mu^+ \mu^-$, which in the SM is produced by single-$W$/$Z$
exchange diagrams.  With respect to the effective Lagrangian, it
corresponds to the four-fermion effective operator
\begin{align}
 &
 \mathcal L_\text{eff:trident} \notag \\
 &
 = \left[ \bar\mu \gamma^\mu \left( C_V - C_A \gamma^5 \right) \mu \right] \left[ \bar\nu \gamma_\mu (1-\gamma^5) \nu \right]\,,  
 \label{EQ:effectiveop}
\end{align}
with an external photon coupling to $\mu^+$ or $\mu^-$. In the SM, we
have $C_V^\text{SM} \neq C_A^\text{SM}$ in Eq.~\eqref{EQ:effectiveop}.
Combining both $W$- and $Z$-exchange diagrams, we
have~\cite{Koike:1971tu,Koike:1971vg,Belusevic:1987cw,Brown:1973ih}
\bea
 C_V^\text{SM} &=& - {g^2 \over 8 m_W^2} \left( {1 \over 2} + 2 \sin^2\theta_W \right) \,, \nn\\
 C_A^\text{SM} &=& - {g^2 \over 8 m_W^2} \, {1 \over 2} \,.  
\eea
On the other hand, the $Z'$ boson contributes to 
Eq.~\eqref{EQ:effectiveop} with the pure $V-A$ form: 
\begin{align}
 C_V^{\rm NP} = C_A^{\rm NP}  = -{g_l^2 \over 4m_{Z'}^2} \,. 
\end{align}

In terms of the coefficients $C_V$ and $C_A$, the inclusive
cross section is given by\footnote{The interference term $C_V C_A$ is
  omitted in Eq.~\eqref{EQ:csresult}.  According to the study in
  Ref.~\cite{Brown:1973ih}, this term is suppressed by an an order of
  magnitude compared to the $(C_{V,A})^2$ terms.}
\cite{Altmannshofer:2014pba}
\begin{align}
 \sigma (\hat s) \simeq \left( C_V^2 + C_A^2 \right) {2\alpha_\text{EM}\, \hat s \over 9\pi^2} \left[ \log\left( {\hat s \over m_\mu^2} \right) - {19 \over 6} \right] \,, 
 \label{EQ:csresult}
\end{align}
for $\hat s = (p_\nu +p_\gamma)^2$, where $p_\nu$ and $p_\gamma$ are
the initial momenta of the neutrino and photon, respectively.  The
existing experimental result~\cite{Mishra:1991bv} for $\sigma (\nu N
\to \nu N \mu^+ \mu^-)$ is compared with $\int \sigma (\hat s) P(\hat
s, q^2)$, where $P(\hat s, q^2)$ is the probability of creating a
virtual photon in the Coulomb field of the nucleus (for example, see
Ref.~\cite{Altmannshofer:2014pba}).  Alternatively, we can compare the
ratio of the experimental data and the SM prediction reported
as~\cite{AGPQ,Altmannshofer:2014pba}
\begin{align} 
 \left. { \sigma_\text{exp.} \over \sigma_\text{SM} } \right|_{\nu N \to \nu N \mu^+ \mu^-} = 0.82 \pm 0.28 \,, 
\end{align}
with the theoretical prediction
%$\sigma_\text{SM+NP} (\hat s) /\sigma_\text{SM} (\hat s)$
%
\begin{align} 
 &
 \left. { \sigma_\text{SM+NP} \over \sigma_\text{SM} } \right|_{\nu N \to \nu N \mu^+ \mu^-}  \\
 &
 \simeq {\sigma_\text{SM+NP} (\hat s) \over \sigma_\text{SM} (\hat s)} 
 =
 { (C_V^\text{SM} + C_V^\text{NP})^2 + (C_A^\text{SM} + C_A^\text{NP})^2 \over (C_V^\text{SM})^2 + (C_A^\text{SM})^2 } \,. \notag
\end{align}
The net effect is that this will provide an upper limit on $g_l^2/m_{Z'}^2$.

\begin{figure}[t]
\hspace{-0.4cm}
\centerline{\includegraphics[width=0.9\hsize]{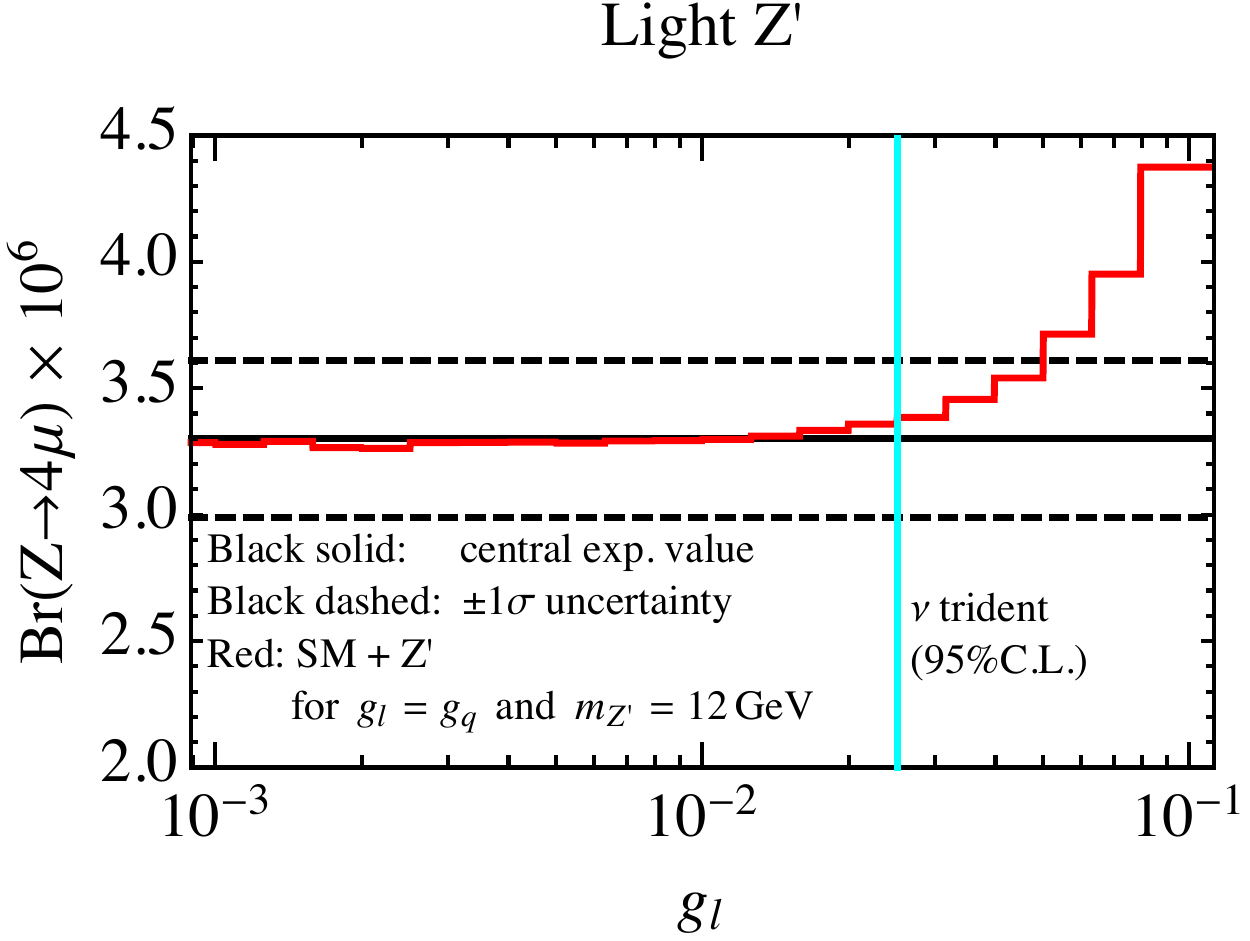}}
\caption{Solid (red): predicted branching ratio for $Z\to 4\mu$ via
  $Z\to Z'\mu^+\mu^-$ for light $Z'$, $m_{Z'}=12\,$GeV, versus $g_l$.
  Horizontal lines denote the $1\sigma$ experimentally-allowed region.
Vertical line is upper limit from $\nu$ trident production.
}
\label{Z4mufig}
\end{figure}

\subsection{\boldmath $Z\to 4\mu$}

A constraint similar to that from neutrino trident production comes
from the process $Z\to \mu\mu^*$, $\mu^*\to \mu {Z'}^*$, ${Z'}^*\to
\mu\mu$, resulting in $Z\to 4\mu$.  The decay mode into light leptons
$(e,\mu)$ has been measured by ATLAS and CMS, giving a branching ratio
consistent with the SM value, $3.3\times 10^{-6}$
\cite{Olive:2016xmw}. The NP contribution is suppressed for heavy
$Z'$, $m_{Z'}> m_Z$, giving a weak constraint, but is larger when
$m_{Z'}<m_Z$ so that the intermediate $Z'$ can be on-shell.  In this
case we can estimate the NP contribution (ignoring interference with
the SM) as
\bea
	\Gamma(Z\to 4\mu) = \Gamma(Z\to Z'\mu^+\mu^-)\, B(Z'\to
\mu^+\mu^-) ~.
\label{Z4mu}
\eea

In our later fit to the AMS-02 antiproton excess, we will be
interested in $m_{Z'} \cong 12\,$GeV.  The predicted branching ratio 
(evaluated with the use of {\tt MadGraph 5}
 \cite{Alwall:2011uj,Alwall:2014hca}) 
is shown in Fig.\ \ref{Z4mufig} for this case, giving the constraint
$g_l < 0.05$.  The result depends upon $g_q$ since this affects
the branching ratio of $Z'\to\mu^+\mu^-$, 
\bea
	B(Z'\to \mu^+\mu^-) &=& {g_l^2\over 2 g_l^2 + 1.9 g_q^2} ~,
\eea
taking account of the phase-space and amplitude suppression for
decays into $b\bar b$.  For definiteness we have taken $g_q=g_l$;
larger values of $g_q$ will weaken the constraint on $g_l$.  Our
result is consistent with the limits obtained in Refs.\
\cite{Altmannshofer:2014pba,Altmannshofer:2014cfa}.

The constraint from $Z\to 4\mu$ is relatively weak; in the case
$g_q=g_l$, the maximum value of $g_l$ consistent with neutrino trident
production (see Fig.\ \ref{flavor}) is $g_l \cong 2 m_{Z'}$/TeV $\cong
0.02$, which is more stringent than that from $Z\to 4\mu$.

\subsection{Muon $\mathbf g-2$}

There has been a long-standing $3.6\sigma$ discrepancy between the
predicted and measured values of the anomalous magnetic moment of the
muon, $a_\mu$. To address this, models have been proposed that include
a $Z'$ with off-diagonal vectorial couplings to $\mu$ and a heavier
lepton ($\ell$). (The case where $\ell$ is a new lepton $L$ is
discussed in Ref.~\cite{Allanach:2015gkd}; $\ell=\tau$ is examined in
Ref.~\cite{Altmannshofer:2016brv}.) This leads to a
$(m_\ell/m_{Z'})^2$ enhancement of the loop contribution to $a_\mu$.

In the present model, the $Z'$ couples only to $\mu$ (and has $V-A$
couplings). The contribution to $a_\mu$ now increases the discrepancy,
though its actual size is too small to be relevant. For example, in
our model with $m_{Z'} =12\,\text{GeV}$, the contribution to $a_\mu$
is negligible as long as $g_l \lesssim 0.02$. And it does not help to
allow the $Z'$ to couple to both $\mu$ and $\tau$ (with off-diagonal
$\mu$-$\tau$ couplings). In this case, there is then a tree-level $Z'$
contribution to $\tau\to 3\mu$, which is strongly constrainted.

%
%Models similar to ours have been studied in connection with the
%long-standing discrepancy between the predicted and observed
%anomalous magnetic moment $a_\mu$ of the muon.  However 
%the NP contribution from $Z'$ exchange in a loop is too small to 
%be relevant unless the $Z'$ has off-diagonal couplings to $\mu$ 
%and a heavier lepton with mass $m_f$, leading to a $(m_f/m_{Z'})^2$ 
%enhancement of the loop contribution to $a_\mu$ \cite{Allanach:2015gkd,Altmannshofer:2016brv}.  
%This is not required in our model for explaining $R_K$, so we do
%not try to address the muon $g-2$ here.

\subsection{Allowed parameter space}

The preceding flavor constraints are summarized in Table~\ref{tab:FC},
where $V_{tb} V_{ts}^* = -0.0405 \pm 0.0012$~\cite{Olive:2016xmw} and
$f_{B_s}\hat B_{B_s}^{1/2} = (266 \pm 18)$ MeV~\cite{Aoki:2016frl}
have been used, and where $\hat m_{\rm TeV} \equiv m_{Z'} /
1\,\text{TeV}$. Concerning $\bs$-$\bsbar$ mixing, the experimental
value is precisely determined (of order $0.1$\%) while the theory
prediction has a large uncertainty.  We take a $1\sigma$ range for the
theoretical uncertainty to obtain the constraint.

%%%%%%%%%%%%%%% [Table] %%%%%%%%%%%%%%%
\begin{table*}[t] 
\renewcommand{\arraystretch}{2}
\begin{tabular}{ccc}
\hline\hline 
 Process					&  Constraint  & Range \\[0.3em]
\hline
 $\bsmumu$				&  $0.00028 \le g_q\, g_l\, s_\theta c_\theta \, \hat m_{\rm TeV}^{-2} \le 0.00177$  & ``$3\sigma$''~\cite{BK*mumulatestfit1} \\[0.3em]
\hline
 $\bsnunubar$				&  $\left| 0.01041 + g_q\, g_l\, s_\theta c_\theta \, \hat m_{\rm TeV}^{-2} \right| \lesssim 0.03711$  &  90\% C.L. \\[0.3em]
\hline
 $\bs$-$\bsbar$ mixing				&  $g_q^2 (s_\theta c_\theta)^2 \, \hat m_{\rm TeV}^{-2} \lesssim 0.00002$  & ($1\sigma$ theor. error)  \\[0.3em]
\hline
% $\nu N \to \nu N \mu^+ \mu^-$~~~~	&  $\left( 16.495 + g_l^2 \, \hat m_{\rm TeV}^{-2} \right)^2 +\left( 31.1853 + g_l^2 \, \hat m_{\rm TeV}^{-2} \right)^2 \le 48.2858$ & 95\% C.L. \\[0.3em]
 $\nu N \to \nu N \mu^+ \mu^-$~~~~	&  $g_l^2 \, \hat m_{\rm TeV}^{-2}\, (1+0.02097 \times g_l^2 \, \hat m_{\rm TeV}^{-2}) \le 4.81193$ & 95\% C.L. \\[0.3em]
\hline\hline 
\end{tabular}
\caption{Summary of the flavor constraints from $\bsmumu$,
  $\bsnunubar$, $\bs$-$\bsbar$ mixing, and $\nu N \to \nu N \mu^+
  \mu^-$, where $\hat m_{\rm TeV} \equiv m_{Z'} / 1\,\text{TeV}$ and
  $s_\theta c_\theta = \sin\theta_D \cos\theta_D$.  
  }
\label{tab:FC}
\end{table*}

\begin{figure}[t]
\hspace{-0.4cm}
\centerline{\includegraphics[width=0.9\hsize]{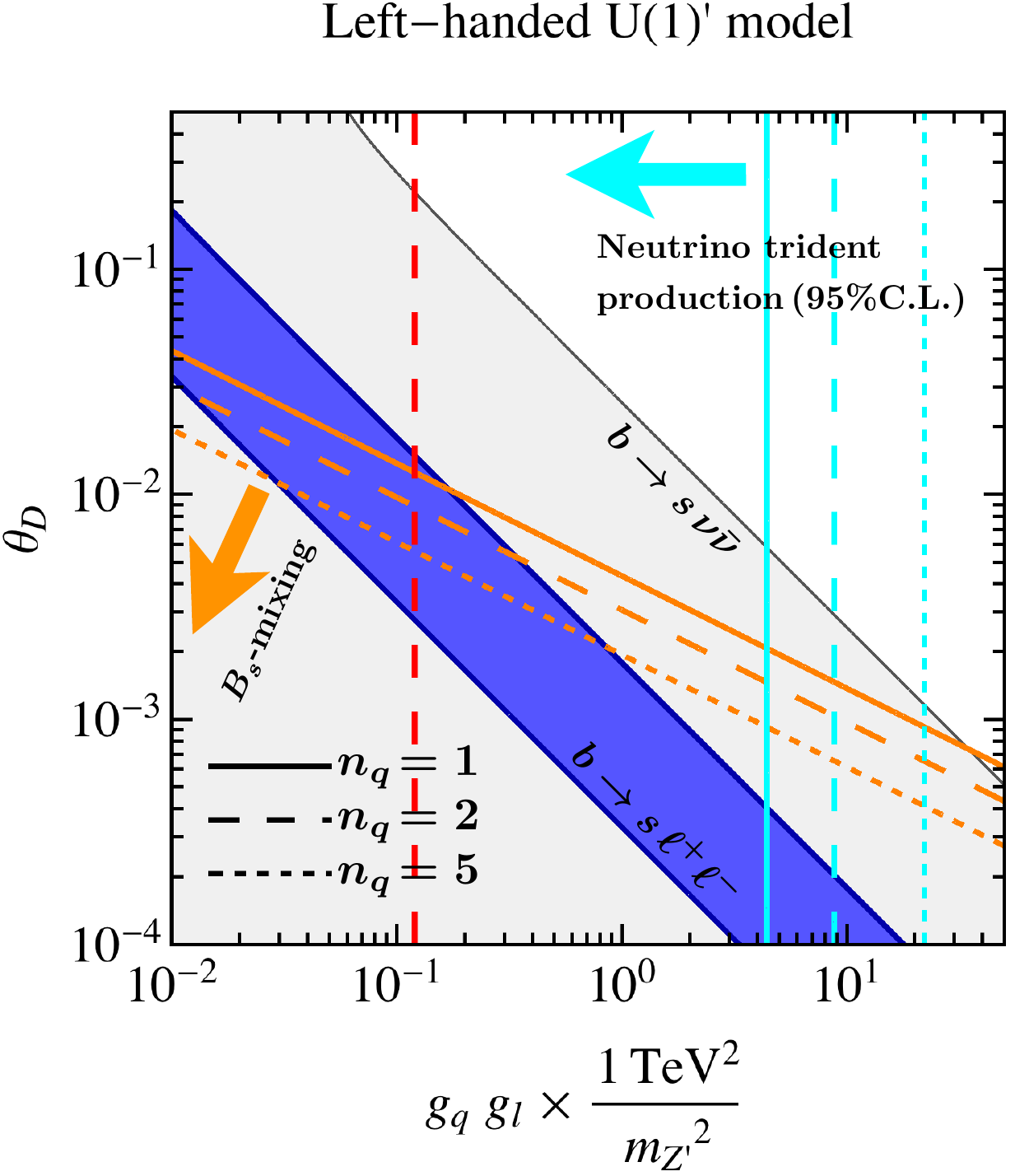}}
\caption{Allowed regions from flavor constraints, for several values of
  $n_q \equiv g_q/g_l$; dark (blue) band gives observed $R_K$.
The preferred couplings from dark matter
  constraints (for the heavy $Z'$ model) are shown by the vertical red
  dashed line (from the $n_q=2$, $n_\chi=5$ model, where $n_q =
  g_q/g_l$ and $n_\chi = g_\chi/g_q$.)  }
\label{flavor}
\end{figure}

In Fig.~\ref{flavor}, we combine all the constraints to determine the
space of allowed values of the theoretical parameters 
in the $(g_qg_l\,\hat m_{\rm
  TeV}^{-2}, \theta_D)$ plane, for several values of $n_q \equiv
g_q/g_l$.  The area in the dark (blue) region below the $B_s$ mixing
lines (orange) and to the left of the neutrino trident lines (cyan)
can explain the $\bsmumu$ anomalies, consistent with all the other
constraints.

Note that Fig.~\ref{flavor} applies for $m_{Z'} \gg m_b$. However, for
the light-$Z'$ scenario ($m_{Z'} = 12$ GeV), the parameter $g_q g_l /
m_{Z'}^2$ should be (approximately) replaced by $g_q g_l / (m_{Z'}^2 -
m_b^2)$.

%%%%%%%%%%%%%%% [Figure] %%%%%%%%%%%%%%%
%%%%%%%%%%%%%%% [Figure] %%%%%%%%%%%%%%%

\section{Dark matter models}
\label{DMsect}

There are two independent tentative anomalies in the AMS-02 antiproton
spectrum: one at low $\sim 10$ GeV energies and one at $\sim 300$ GeV.
To alternatively address them, we consider two possible extensions of
the model to include dark matter: (1) TeV-scale $Z'$, and Dirac dark
matter of mass 30-70 GeV, and (2) 10 GeV-scale $Z'$, coupled to two
quasi-degenerate Majorana DM states with masses $m_\chi \sim 2\,$TeV.
In the second model, the $Z'$ couples off-diagonally to the DM mass
eigenstates, alleviating direct detection signals.  A consistent
treatment of the second model requires the inclusion of the dark Higgs
boson that gives mass to the $Z'$.

\subsection{Heavy $Z'$, Dirac dark matter}

We first consider the scenario in which the DM $\chi$ is a Dirac
particle with mass $m_\chi\ll m_{Z'}$ and vectorial coupling to the
$Z'$ with strength $g_\chi$.  In the approximation of small mixing
angles, where we neglect the couplings to lower-generation quarks, the
$Z'$ can be integrated out to give the effective Hamiltonian
\bea
	H &=&
	{g_q\, g_\chi\over m_{Z'}^2}\sum_{i=t,b}
	(\bar q_i\gamma_\mu P_\L q_i)
	(\bar \chi\gamma^\mu \chi)\nn\\
	&& ~+ 
	{g_l\, g_\chi\over m_{Z'}^2}\sum_{j=\mu,\nu_\mu}
	(\bar l_j\gamma_\mu P_\L l_j)
	(\bar \chi\gamma^\mu \chi) ~.
\label{Ham}
\eea
As in the preceding sections, we assume that the $Z'$ couples only to
left-handed SM particles.

\subsubsection{Astrophysical constraints}
 
The cross section for $\chi\bar\chi$ annihilation into $b_\L$ quarks and
$\mu_\L, \nu_\mu$ leptons is given by
\be
	\langle\sigma v\rangle = {(3g_q^2 + 2g_l^2)\, m_\chi^2\over 2\pi}
	\left({g_\chi\over m_{Z'}^2}
	\right)^2 \cong 4.4\times 10^{-26}{\rm\,cm^3\over s}
\label{relic}
\ee
to get the right relic density \cite{Steigman:2012nb}.  This is the
appropriate formula for $m_\chi < m_t$, as suggested by the best-fit
regions for AMS excess antiprotons, $m_\chi \in [30-70]\,$GeV
\cite{Cui:2016ppb}, or $m_\chi\cong 80\,$GeV
\cite{Cuoco:2016eej}.\footnote{\label{fn3} Ref.\ \cite{Lin:2016ezz}
  finds a larger DM mass of $m_\chi\cong 200\,$GeV as the best-fit
  point, which would give a larger predicted cross section, with
  $(3g_q^2 + 3g_l^2)\to (10.1\,g_q^2 + 3g_l^2)$, due to the production
  of top quark pairs with some phase-space suppression
  ($(1-m_t^2/m_\chi^2)^{1/2}$), compensated by matrix element
  enhancement $(1 + m_t^2/2 m_\chi^2)$.  We find this scenario is
  difficult to reconcile with the global constraints, and hence do not
  further consider it.}

To get a large enough antiproton signal, consistent with the thermal
relic annihilation cross section, we want quarks to dominate in the
final state.  Reducing the relative coupling to leptons also helps to
alleviate stringent LHC constraints considered below, but at the same
time diminishes the NP contribution to $\bsmumu$.  We find that taking
$g_q = n_q\, g_l$ with $n_q=2$ is a sufficient compromise, implying
that annihilation into $b$ quarks makes up 86\% of the total cross
section.  This leaves just one ratio $g_\chi/g_q \equiv n_\chi$  to be constrained.
We then have from Eq.~(\ref{relic})
\bea
    g_\chi &=& {1.09\sqrt{n_\chi}\over \hat m_{70}^{1/2}}\,{m_{Z'}\over {\rm TeV}} ~, \nn\\
    g_q = 2 g_l &=& {1.09\over \sqrt{n_\chi}\,\hat m_{70}^{1/2}}\,{m_{Z'}\over {\rm TeV}} ~,
\label{couplings}
\eea
where $\hat m_{70} \equiv m_\chi/(70\,{\rm GeV})$.

The couplings in (\ref{couplings}) are evaluated at the scale of
$m_\chi$, after integrating out the heavy $Z'$ at the scale of its
mass.  It has been pointed out in Ref.\ \cite{Davoudiasl:2015hxa} that running of
the U(1)$'$ coupling in dark matter models can sometimes be
important. However below the $Z'$ threshold, no significant running
is expected because the $Z'$ is heavy and has already been removed
from the effective theory.  We have estimated this effect by computing
the vertex correction with loop momenta between $m_\chi$ and $m_{Z'}$
with a massive $Z'$ propagator,
finding that $\Delta g_\chi\sim 3\times 10^{-3} g_\chi^3$.  
 On the other hand, above $m_{Z'}$ 
running can become significant, and one can wonder whether
perturbation
theory may break down at scales not far above $m_{Z'}$.

\begin{figure}[t]
\hspace{-0.4cm}
\centerline{\includegraphics[width=\hsize]{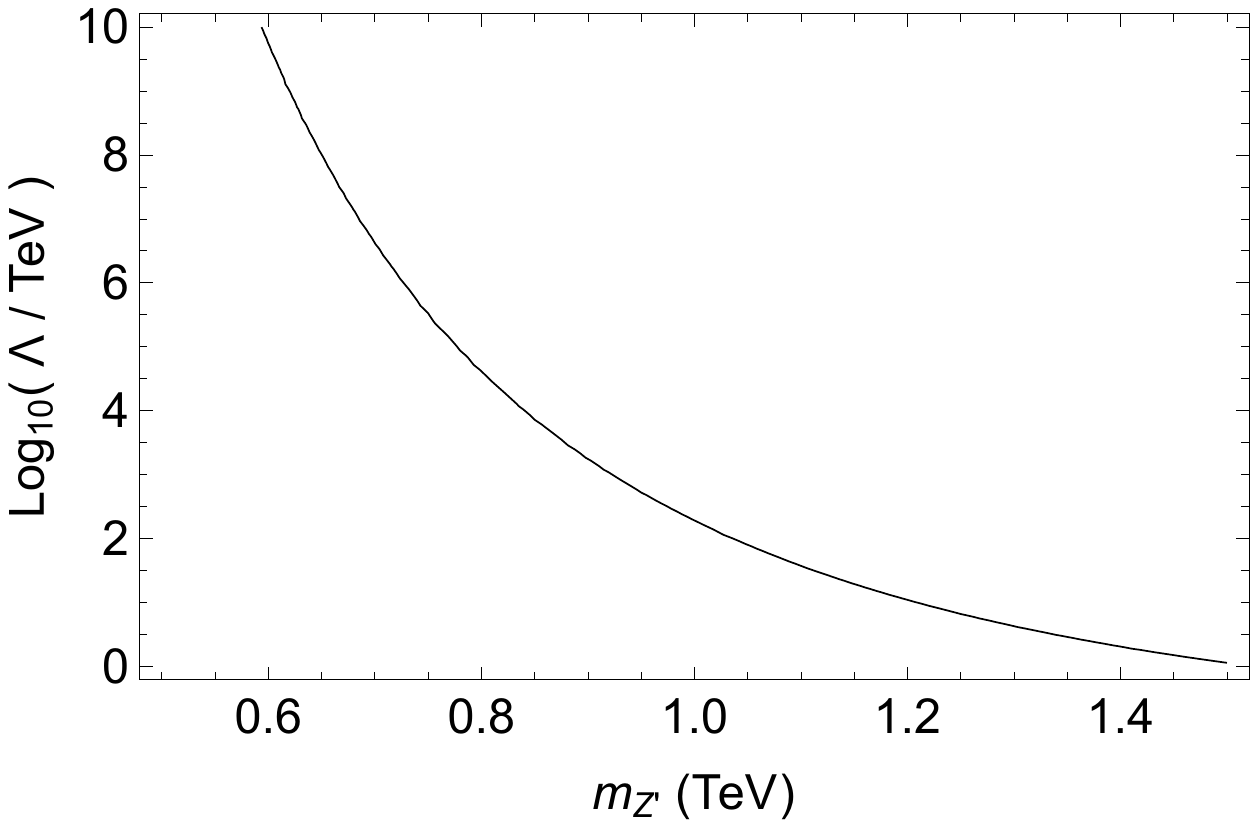}}
\caption{The UV scale $\Lambda$ where $\alpha_\chi = g_\chi^2/4\pi =1$, 
versus $m_{Z'}$, using the relic density value of
$g_\chi$ from (\ref{couplings}) at the scale $m_\chi = 70\,$GeV as the
IR boundary condition for RG running.}
\label{running}
\end{figure}

 To estimate whether this is the case in the present
model, we integrate the beta function $d g_\chi/d\ln\mu = 
g_\chi^3/12\pi^2$ between $m_{Z'}$ and a UV scale $\Lambda$, 
finding that
$\alpha_\chi = g_\chi^2/4\pi\gtrsim 1$ already at $\Lambda = 10\,$TeV
for $m_{Z'}= 1.2\,$TeV, while for smaller $m_{Z'}$ the scale
of nonperturbativity quickly rises to much higher values, as shown in
fig.\ \ref{running}.
Therefore lighter $m_{Z'} \lesssim 1\,$TeV are preferred for 
consistency of the theory up to scales above $\sim 100\,$TeV,
where some UV completion could be expected.

The most recent Fermi-LAT searches for emission from dark matter
annihilation in dwarf spheroidal galaxies currently exclude cross
sections of $\langle \sigma v \rangle > 1.9 \times 10^{-26}$ cm$^3$/s
at 95\% C.L. for 80 GeV DM annihilating to $b \bar b$
\cite{Ackermann:2015zua}. This is in tension with the cross sections
suggested by the DM interpretation of the $\bar p$ excess. However,
recent works \cite{Bonnivard:2015xpq,Ullio:2016kvy} have pointed out
that the dark matter content of some of the dwarf spheroidals in the
Fermi analysis may have been overestimated, resulting in a less
stringent limit that can be compatible with DM explanations of cosmic
ray excesses.

\begin{figure*}[t]
\hspace{-0.4cm}
\centerline{\includegraphics[width=0.45\hsize]{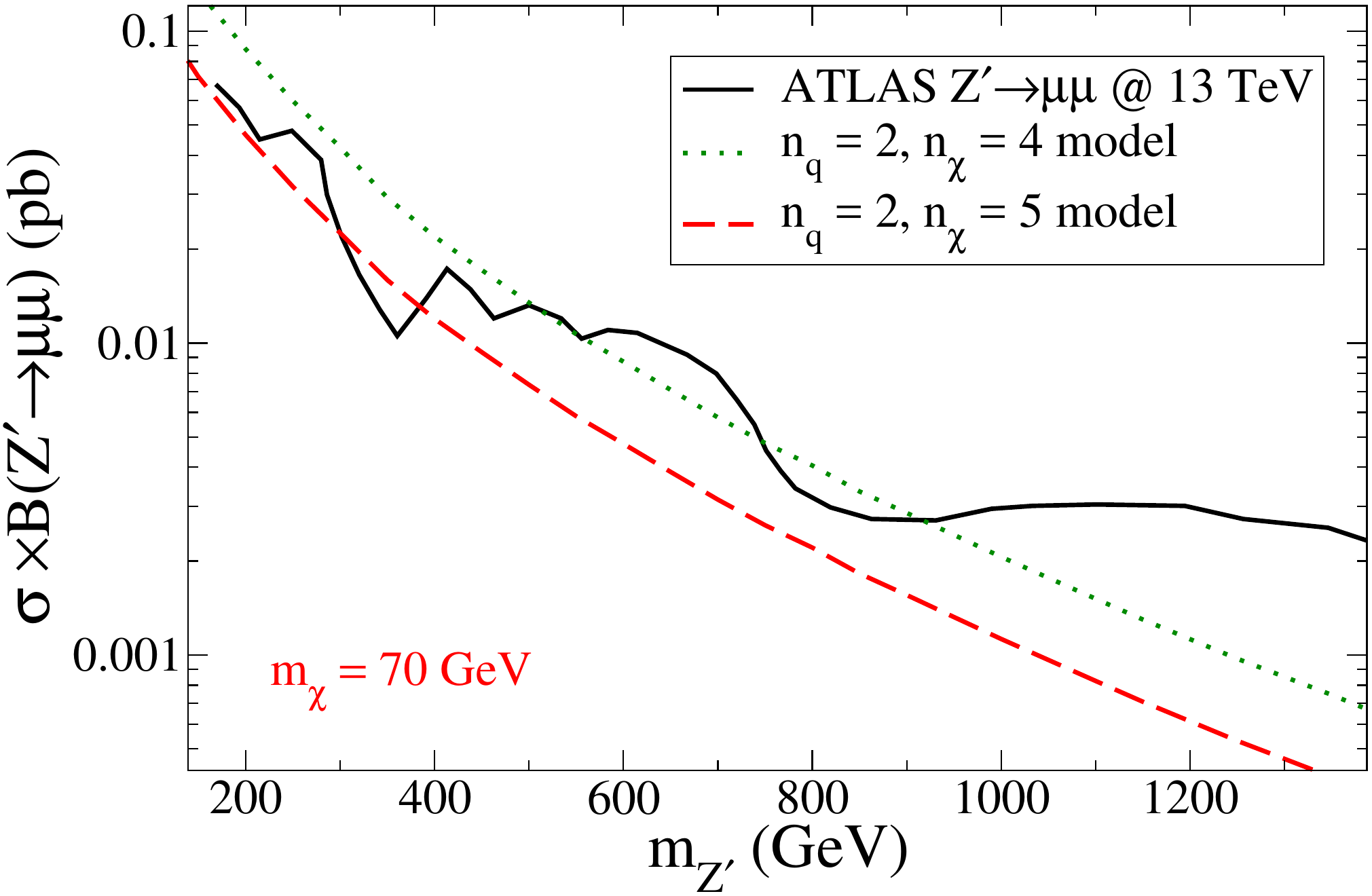}
\hfil\includegraphics[width=0.45\hsize]{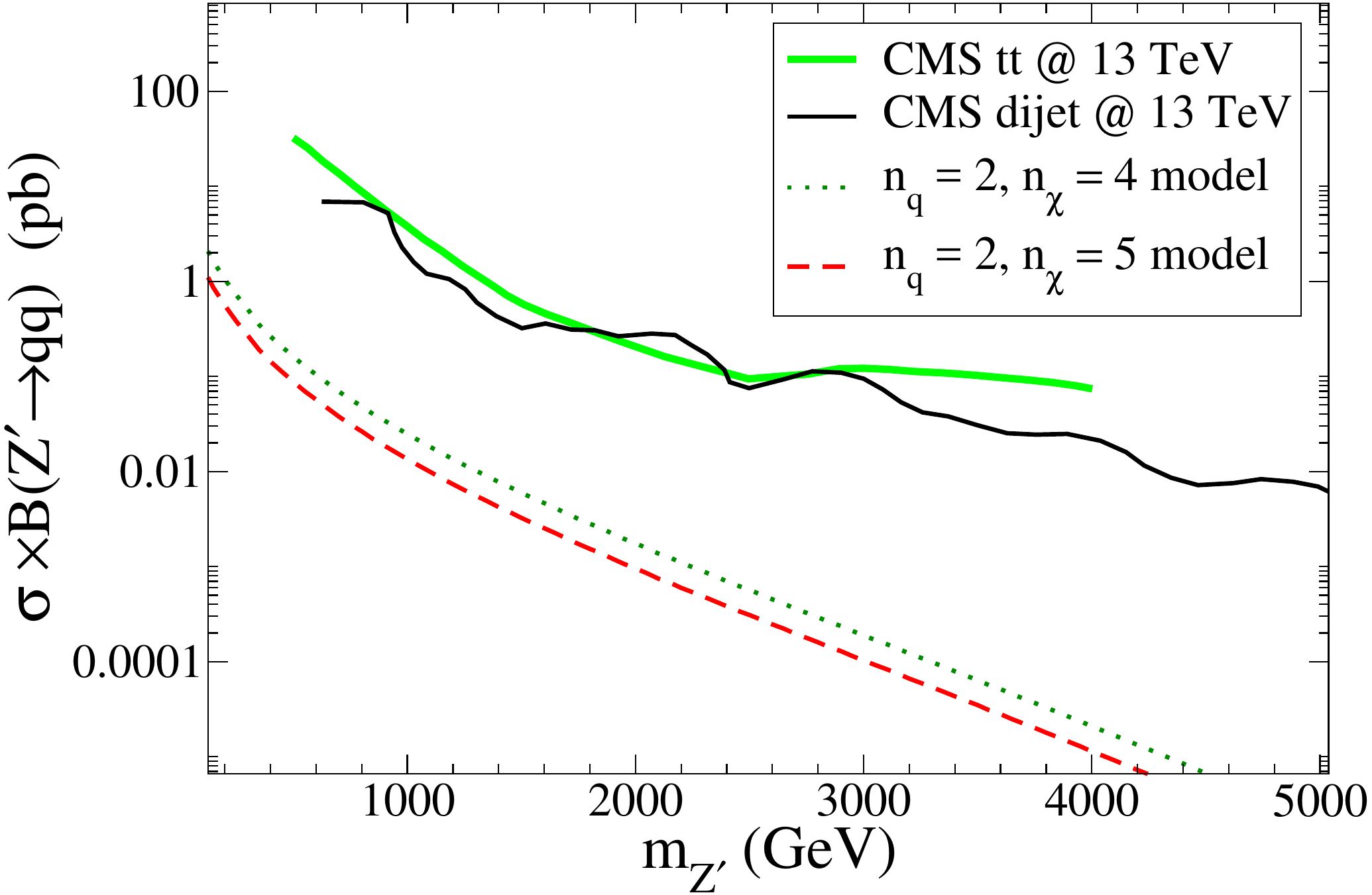}}
\caption{Left: ATLAS limit on $pp\to Z'\to\mu\bar\mu$ production and
  decay, and predictions of two models that
  are close to the constraint; right: same for $pp\to Z'\to b\bar b$
  or $t\bar t$ as limited by searches for dijet or $t\bar t$ final
  states. The dijet limit is adjusted upward from the published value
  of $\sigma B_{qq} A$ by assuming the event acceptance is $A=0.6$
  \cite{Sirunyan:2016iap}.}
\label{limit-mumu}
\end{figure*}

\subsubsection{Collider limits}

ATLAS and CMS have searched for resonant lepton pairs from
$Z'\to\ell\bar\ell$ \cite{ATLAS:2016cyf,CMS:2016abv}.  These depend on
the branching ratio of $Z'$ into $\mu^+\mu^-$, which in our model is
given by
\bea
	B(\mu\bar\mu) &=& {g_l^2\over 3(1+f)\,g_q^2 + 2\,
	g_l^2 + 2\,g_\chi^2} \nn\\
	&=& {0.25\over 3.5 + 3f + 2 n_\chi^2} ~,
\label{BR}
\eea
where $f=(1 + 7x/17)\sqrt{1-4 x^2}$ with $x=(m_t/m_{Z'})^2$ for top
quark final states \cite{Cline:2014dwa}.
It is common in model-building to forbid $Z'$ couplings to leptons
in order to avoid these stringent dilepton constraints.  Here we
manage to satisfy them by coupling the $Z'$ only to the $b$-quarks
present in the proton, leading to PDF suppression of the production
cross section, combined with a reduction in the partial width of  
$Z'$ to leptons due to the invisible decays $Z'\to\chi\bar\chi$.

We show the ATLAS dilepton limit in Fig.\ \ref{limit-mumu} (left),
along with predictions for the model with $g_l = g_q/n_q = 0.5\,g_q$,
$g_\chi = n_\chi\,g_q = 5\,g_q$, and $m_\chi = 70\,$GeV, for which the
region with $300\,$GeV $< m_{Z'} < 390\,$GeV is excluded. These were
calculated by computing the production cross section for $Z'$ 
through its coupling to $b$-quarks using {\tt MadGraph 5} 
 \cite{Alwall:2011uj,Alwall:2014hca}, with a 
QCD $K$-factor correction that happens to be unity within
uncertainties of $\sim 10-30\%$ (see fig.\ 3 of Ref.\ 
\cite{Faroughy:2016osc}). Then
eq.\ (\ref{couplings}) implies
$g_l g_q/m_{Z'}^2 = 0.12/$TeV$^2$, which is shown as
the vertical line in the parameter space relevant for $b\to
s\mu^+\mu^-$, Fig.\ \ref{flavor}.  The blue region below the dashed
lines, showing the upper bound on the quark mixing angle from $B_s$
mixing, is allowed.

Eq.\ (\ref{couplings}) demands a large coupling $g_\chi$ unless
$m_{Z'}$ is in the lower part of its allowed region.  For example,
with $m_{Z'} = 250\,$GeV, we obtain $g_\chi = 0.6$ (and it scales
linearly with $m_{Z'}$ for larger values).  Taking larger values of
$m_\chi$ reduces the couplings needed to get the right relic density,
and further alleviates tension with the dilepton search, but it also
pushes $g_q g_l/m_{Z'}^2$ further to the left in Fig.\ \ref{flavor},
making it difficult to get a large enough contribution to $\bsmumu$.
This is the problem with the scenario with $m_\chi = 200$\,GeV (see
footnote \ref{fn3}).

There are also limits from resonant dijet searches from $b\bar b$ or
$t\bar t$ final states
\cite{Aaboud:2016nbq,ATLAS:2016gvq,CMS:2016zte,Sirunyan:2016iap} but
which are weaker than those from the dilepton searches.  The branching
ratio to $b$ quarks is 12 times greater than Eq.~(\ref{BR}), but the
predicted cross section is still far below the limit, as shown in
Fig.\ \ref{limit-mumu} (right).

\begin{table*}
\centering
\begin{tabular}{c||c|c||c|c}
\hline \hline
\textbf{Propagation} & \multicolumn{2}{c||}{$\chi \chi \to q \bar q$
\, $m_{Z'} = 5$ GeV}  & \multicolumn{2}{c}{$\chi \chi \to b \bar b$
\, $m_{Z'} = 12$ GeV} \\
\textbf{Model}& $m_\chi$ [GeV] & $\langle \sigma v \rangle$ [$10^{-26}$ cm$^3$/s]& $m_\chi$ [GeV] & $\langle \sigma v \rangle$ [$10^{-26}$ cm$^3$/s] \\
\hline
MIN & 765 & $18.6{+10.7\atop -8.0}$ & 1800 & $103{+59\atop -44}$ \\
MED & 808 & $5.2{+3.0\atop -2.4}$   & 1950 & $31{+18\atop -14}$  \\
MAX & 826 & $2.29{+1.3\atop -1.1}$  & 1950 & $12.8{+7.3\atop -5.9}$ \\
\hline \hline
\end{tabular}
\caption{\label{xsecs} The values on the left are the best-fit values
  of dark matter mass and self-annihilation cross section for
  explaining the $\bar p$ excess as determined in
  Ref.~\cite{AMSexplain}. These fits were done considering mediators of
  mass 5 GeV which decay to light quarks ($q = u,d$) for the three
  standard propagation parameter sets. On the right are the values of
  $m_\chi$ that give roughly the same prompt spectrum of $\bar p$ when
  the mediator has a mass of 12 GeV and decays exclusively to $b$
  quarks (see Fig.\ \ref{boosted}).  Also listed are necessary cross
  sections to achieve the same dark matter annihilation rate for these
  masses.}
\end{table*}

\subsubsection{Direct detection}

The couplings of $Z'$ to light quarks in this model are highly
suppressed, making the tree-level contribution to $\chi$-nucleon
scattering well below the current limit.
  The coupling of $Z'$ to left-handed up quarks due
to mixing is of order [see Eq.~(\ref{Z'uu})]
\be
	g_u \sim |\theta_D V_{us} - V_{ub}|^2 g_q \sim 6\times 10^{-6} g_q 
\label{ucoupling}
\ee
for the maximal quark mixing angle $\theta_D = 0.008$ indicated in
Fig.\ \ref{flavor}.  The effective cross section on nucleons is given
by\footnote{We correct an erroneous factor of 4 in their formula}
\cite{Arcadi:2013qia}
\be
	\sigma_N = {(g_\chi g_u m_n)^2\over 4\pi\, m_{Z'}^4}
	\left(1 + Z/A\right)^2 \cong {2\times 10^{-51}{\rm\, cm}^2
	} ~,
\ee
using Eqs.~(\ref{couplings},\ref{ucoupling}) with $m_\chi = 70\,$GeV,
where $m_n$ is the nucleon mass.  This is well below the expected
reach of the LZ experiment, $2\times 10^{-48}{\rm\, cm}^2$
\cite{Akerib:2015cja}.

However, the coupling of $Z'$ to quarks and leptons contributes at one
loop to kinetic mixing, $(\epsilon/2) F^{\mu\nu} Z'_{\mu\nu}$.  The
contributions are logarithmically divergent, and only cancel if $g_q =
g_l$.  To estimate the natural size of such corrections in the model
with $g_q = 2g_l$, we imagine that there is some heavy vector-like
fermion with mass $m_F$ and charges such that it cancels the UV
contributions of the SM fermions to $\epsilon$ at scales above $m_F$.
Then, in the infrared one finds
\be
	\epsilon \cong {e g_q\over 24\pi^2}\ln\left({m_t^4 \over m_b^2 m_\mu m_F}
	\right) \sim 0.036\, g_q e ~,
\label{kmix}
\ee
where we have taken $m_F = 100\,$ TeV to get the numerical estimate.
This provides an example of how loop effects from the coupling of
new physics to leptons (in this case $\mu$) can be important for
the coupling to quarks relevant for direct detection, as has been
discussed with respect to leptophilic dark matter models in 
Ref.\ \cite{DEramo:2017zqw}.

Kinetic mixing leads to the effective interaction 
\be
	{\epsilon e g_\chi\over m_{Z'}^2}(\bar\chi\gamma^\mu\chi)
	(\bar p\gamma_\mu p)
\ee
between DM and protons.  The cross section on protons is then
\be
	\sigma_p = {(\epsilon e g_\chi m_p)^2\over \pi m_{Z'}^4} \sim 
	{1.7\times 10^{-45}\over \hat m_{70}^2}{\rm\, cm}^2 ~,
\label{sigmap}
\ee
where $m_p$ is the proton mass, and we have used
Eqs.\ (\ref{couplings},\ref{kmix}).  This is just below
the current limit of $1.8\times 10^{-45}{\rm\, cm}^2$ on protons
for 70 GeV DM from the
PandaX-II experiment \cite{Yang:2016odq}, and well above the expected
reach of LZ experiment, $1\times 10^{-47}{\rm\, cm}^2$ for DM coupling
to protons.

\begin{figure}[t]
\hspace{-0.4cm}
\centerline{\includegraphics[width=0.9\hsize]{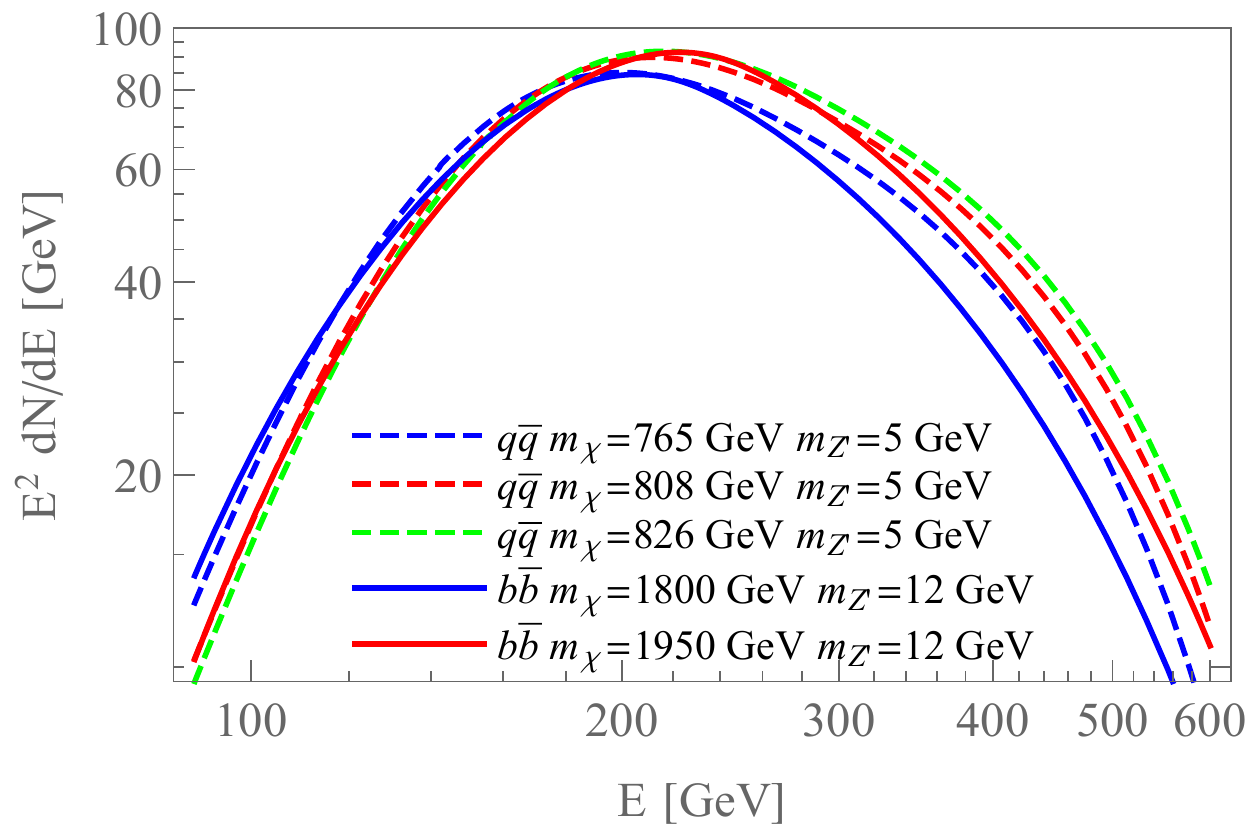}}
\caption{Antiproton spectra from $\chi\chi\to Z'Z'\to b{\bar b} \,
  b{\bar b}$ for $m_{Z'} = 12\,$GeV and several dark matter masses,
  compared to the best fit $\chi\chi\to Z' Z' \to q{\bar q} \, q{\bar
    q}$ spectra found in Ref.\ \cite{AMSexplain}.  }
\label{boosted}
\end{figure}

\subsection{Light $Z'$, Majorana dark matter}

Here we discuss an alternative scenario in which TeV-scale DM
annihilates into highly-boosted light $Z'$ bosons, whose subsequent
decays into $b$ quarks produce antiprotons with a sharply peaked
spectrum, to explain a tentative bump at high energies in the AMS-02
data.

\subsubsection{Antiproton spectrum}

Ref.\ \cite{AMSexplain} recently observed that heavy DM, with
$m_\chi \sim (0.6$\,--\,1)\,TeV, annihilating into light mediators of
mass $\sim 5$\,GeV that decay to $u$ and $d$ quarks, can lead to a
spectrum of $\bar p$ that fits the AMS-02 excess at high energies.
The decay products are highly boosted and result in $\bar p$'s that
have a spectrum peaked near 300\,GeV as observed.  The required
annihilation cross sections, depending upon different models of cosmic
ray propagation, are listed in Table \ref{xsecs}.  These sets of
propagation parameters are not the standard ones that appear in the
literature (e.g., Ref.~\cite{Donato:2003xg}), but rather a more recent
fit to the proton flux and B/C ratio as measured by AMS-02
\cite{Jin:2014ica}.

The best-fit values of $\langle\sigma v\rangle$ show that dark matter
explanations of the excess tend to require an annihilation cross
section above the thermal relic value, $2.3\times 10^{-26}$cm$^3$/s
for 800\,GeV DM \cite{Steigman:2012nb}, suggesting that a complete
model should have a mechanism, such as Sommerfeld enhancement, for
boosting the late-time annihilation cross section relative to that in
the early universe.

%The values of $m_Z'$ and
%$m_\chi$ determine the shape of the prompt antiproton spectrum, which is
%chosen to fit the excess observed by AMS. 

The prompt $\bar p$ spectrum produced by dark matter annihilation in
this scenario is found by boosting the spectrum of $\bar p$ from the
decays of two $Z'$ bosons at rest. It is given by \cite{AMSexplain}
\be
    \frac{dN(x)}{d x} = 2 \int^{b(x)}_{a(x)} dx'
    \frac{1}{\sqrt{1-E_1^2} \sqrt{x'^2 - E_0^2}}
    \frac{dN(x')}{dx'} ~,
\ee
where $x = E/m_\chi$, $E$ is the total energy, $x'=2E'/m_{Z'}$, $E_1 =
m_{Z'}/m_\chi$, and $E_0 = 2 m_{\bar p}/m_{Z'}$. The upper and lower
limits of integration are $a(x) = x_-$ and $b(x) = {\rm min}
\{1,x_+\}$ with $x_\pm = 2 (x \pm \sqrt{(1-E_1^2)(x^2 - E_1^2 E_0^2 /
  4)})/E_1^2 $. Therefore the prompt spectrum of $\bar p$ from a dark
matter annihilation in this model is determined by $m_\chi$, $m_{Z'}$
and the spectrum of $\bar p$ from a single $Z'$ decay. For the latter,
we use the tabulated spectra in the PPPC 4 DM ID \cite{Cirelli:2010xx,
  Ciafaloni:2010ti}.

In Ref.\ \cite{AMSexplain}, it was assumed that $Z'$ decays with
equal strength into light quarks $q=u,d$, whereas in our model, it
decays predominantly to $b$ quarks. We find that, to achieve nearly the
same shape of the spectrum for $Z' \to b \bar b$ as for decays to $q
\bar q$, we require larger values of both the DM and $Z'$ masses, as
shown in Fig.\ \ref{boosted}. For such a light (12 GeV) $Z'$, fits to $b\to
s\mu^+\mu^-$ should be in terms of $g_q g_l/(m_{Z'}^2 - m_b^2)$,
leading to a 12\% reduction in the required size of $g_q g_l$ compared
to the $m_{Z'}\gg m_b$ limit.  More importantly, since the rate of
annihilation in the galaxy scales as $n_\chi^2\langle\sigma v\rangle$
and $n_\chi \sim 1/m_\chi$, we need to increase the target values of
$\langle\sigma v\rangle$ accordingly.  As in the previous section, we
consider $g_q\gtrsim 2\,g_l$ so that decays to leptons can be
ignored. The rescaled cross sections and dark matter masses relevant
for our model are shown in the right side of Table \ref{xsecs}.

Fermi-LAT searches for DM annihilation in dwarf spheroidal galaxies
currently exclude annihilation cross sections of $\langle \sigma v
\rangle > 42 \times 10^{-26}$ cm$^3$/s at 95\% C.L. for 1.95 TeV DM
annihilating to $b \bar b$ \cite{Ackermann:2015zua}, in tension with
the value needed to explain the $\bar p$ excess with the MIN
propagation model. Ref.\ \cite{AMSexplain} has shown that the
tension is ameliorated for the case of interest where $\chi\bar\chi
\to b{\bar b} \, b{\bar b}$.

\subsubsection{Dark Matter Model}

To avoid stringent constraints from direct detection with such a light
mediator, we wish to forbid vector couplings of the $Z'$ to $\chi$.  A
simple model that accomplishes this, while also explaining the origin
of the $Z'$ mass, has the Lagrangian \cite{TuckerSmith:2001hy}
\bea
	{\cal L} &=& \bar\chi\left(i\slashed{\partial} - g_\chi
	\slashed{Z'} - M\right)\chi -\left({f\over\sqrt{2}}\phi \bar\chi \chi^c
	+{\rm h.c.}\right)\nn\\
	&& ~+ \left|(\partial_\mu -2ig_\chi Z'_\mu)\phi\right|^2-
	\lambda'(|\phi|^2 -\sfrac12 w^2)^2,
\eea
where $\chi$ is a Dirac particle and the scalar potential causes
$\phi$ to get a VEV $\langle\phi\rangle \equiv w/\sqrt{2}$.
After symmetry breaking, $\chi$ splits into two Majorana states
$\chi_\pm = {1\over\sqrt{2}}(\chi \pm \chi^c)$, with masses $M_\pm =
M\pm f w$.  The resulting dark sector Lagrangian includes the terms
\bea
	{\cal L} &\ni& \sfrac12\sum_\pm \bar\chi_\pm\left(i\slashed{\partial} 
	 - M_\pm\right)\chi_\pm - {g_\chi\over 2} 
	\left(\bar\chi_+\slashed{Z}'\chi_- + {\rm h.c.}\right)\nn\\
	&& ~- \sfrac12\sum_\pm \pm f\varphi\bar\chi_\pm\chi_\pm 
	+ \frac{1}{2}\left( \partial_\mu \varphi \partial^\mu \varphi
	- m_\varphi^2 \varphi^2\right)
	 \nn \\
    && ~+ \sfrac{1}{2} m_{Z'}^2 Z'_\mu Z'^\mu + 2 g_\chi^2 Z'_\mu Z'^\mu \left(2 w \varphi + \varphi^2 \right) ~,
\eea
where $\varphi$ is a dark Higgs boson defined by 
$\phi = {1\over\sqrt{2}}(w + \varphi)$, $m_\varphi =
(2\lambda')^{1/2}w$, and $m^2_{Z'} = (2 g_\chi w)^2 +
(g_q\langle\Phi_q\rangle)^2 + (g_l\langle\Phi_l\rangle)^2$ .  

Recall that the fields $\Phi_{q,l}$ were introduced
in eq.\ (\ref{eq1}) for generating Yukawa couplings that would otherwise
be forbidden by the U(1)$'$ symmetry.  In order to help keep $m_{Z'}$
sufficiently light, we assume here that $\langle\Phi_l\rangle \ll
\langle\Phi_q\rangle$ so that its contribution to $m_{Z'}$ can be 
neglected.  Moreover we adhere to the relatively small values of 
$g_q = 2 g_l = 0.4\,m_{Z'}$/TeV$= 0.005$ that were preferred in the 
heavy $Z'$ scenario, but now in order to keep 
$g_q\langle\Phi_q\rangle\cong 4.2\,$GeV
sufficiently small (recalling our assumption that $\langle\Phi_q\rangle
\cong M_t\cong 870\,$GeV to obtain the observed top Yukawa coupling).

Using these values, $m_{Z'}$ is generated 
primarily by the first term $2 g_\chi w \cong 11\,$GeV.  We take
these parameter values as an example; it would be possible to choose
somewhat larger $g_{q,l}$, allowing for the $Z'$ to get somewhat more
of its mass from $\langle\Phi_q\rangle$ at the expense of smaller
values of $w$.  It will become apparent that taking too small values of $w$
would violate a technical assumption we make below for simplifying the
analysis of Sommerfeld enhancement in $\chi$ annihilation.

A key feature of this model is that as long as $fw\gtrsim 50\,$keV,
there are no constraints from direct detection since the ground state
$\chi_-$ does not have enough energy to produce $\chi_+$ in an
inelastic collision with a nucleus.  The tree-level decay
$\chi_+\to\chi_-\nu_\mu\bar\nu_\mu$ mediated by a $Z'$ is
kinematically allowed even for such small mass splittings, so in the
present day the dark matter is made up entirely of $\chi_-$.

We note that it would not be natural to make 
$m_\varphi\gg m_{Z'}$ since both are
of order $w$, so a
consistent treatment demands that we include it in the Lagrangian.
Doing so also avoids problems with tree-level unitarity that would
occur in models with axial couplings of light $Z'$ vector bosons to
heavy DM \cite{Kahlhoefer:2015bea}.  In the present case, we will find
that dark Higgs exchange plays an important role by providing a
Sommerfeld enhancement of DM annihilations in the galactic halo.

\subsubsection{Relic density}

The couplings of $\chi_\pm$ to both $Z'$ and $\varphi$ after breaking
of the $U(1)'$ symmetry lead to several annihilation processes that
can affect the DM relic abundance; these include $\chi_\pm \chi_\pm
\to Z' Z'$ and $\chi_+ \chi_- \to Z' \varphi$. Also present is
$\chi_\pm \chi_\pm \to \varphi \varphi$, but it is $p$-wave suppressed
and so we neglect it.  Since the $\bar p$ signal requires $m_{Z'} \ll
m_{\chi_-}$, we expand the cross section in powers of
$m_{Z'}/m_{\chi_-}$ and keep only the leading terms.  As noted above,
the dark Higgs mass cannot be much larger than $m_{Z'}$, so we neglect
terms suppressed by $m_\varphi/m_{\chi_\pm}$. In the kinematic threshold
approximation $v_{\rm rel}\cong 0$, the annihilation cross sections
are
\bea
\label{chimxsect}
    \langle\sigma v\rangle_{\chi_\pm \chi_\pm \to Z' Z'} &\cong& 
        {g_\chi^4 \over 16\pi\, m_{\chi_-}^2}
        \left( 1 - 2 \frac{f m_{Z'}}{g_\chi m_{\chi_-}} \right) ~, \\
    \langle\sigma v\rangle_{\chi_+ \chi_- \to Z' \varphi} &\cong&
        {(g_\chi^2 - f^2)^2 \over 16\pi\, m_{\chi_-}^2}
        \left( 1 - \frac{f m_{Z'}}{g_\chi m_{\chi_-}} \right)  \, .
\eea

\begin{figure}[t]
\hspace{-0.4cm}
\centerline{\includegraphics[width=0.9\hsize]{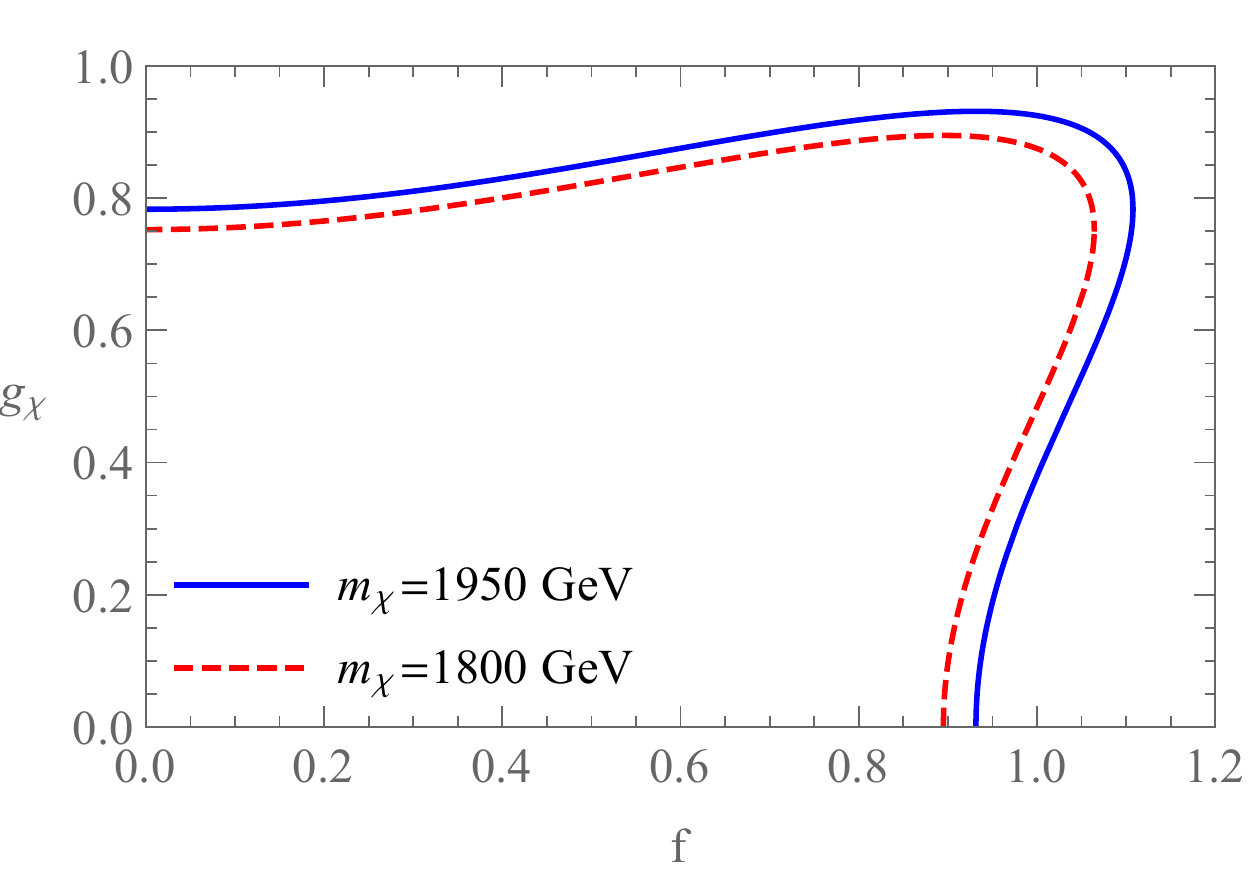}}
\caption{Values of $g_\chi$ and $f$ that give the correct relic
  density for $m_\chi=1950\,$GeV (MED and MAX propagation models) and
  $m_\chi=1800\,$GeV (MIN model).}
\label{fig:relic}
\end{figure}

Both $\delta m_\chi = 2 f w$ and $m_{Z'}\cong 2g_\chi w$ are proportional
to $w$, so the $\chi$ mass splitting must also be $\lesssim 10\,$GeV
(but not so small that inelastic scattering with nuclei becomes
possible). Therefore it is a good approximation to take $m_{\chi_+}
\cong m_{\chi_-}$ in estimating the relic density.  The effective
annihilation cross section in this limit is \cite{Griest:1990kh}
\bea
    \langle\sigma v\rangle_{\rm eff} &=& \sfrac{1}{4}  \langle\sigma v\rangle_{\chi_+ \chi_+ \to Z' Z'}
                         + \sfrac{1}{2}  \langle\sigma v\rangle_{\chi_+ \chi_- \to Z' \varphi} \nn \\
                       && ~+ \sfrac{1}{4}  \langle\sigma v\rangle_{\chi_- \chi_- \to Z' Z'} \, .
\eea
The coefficients for $\chi_\pm \chi_\pm \to Z' Z'$ are half that for
$\chi_\pm \chi_\mp \to Z'\phi$ because the former process has
identical Majorana fermions in the initial state. The correct relic
density in this case requires $\langle\sigma v\rangle_{\rm eff} \cong
2.3\times 10^{-26}{\rm cm}^3/{\rm s}$ \cite{Steigman:2012nb}, giving a
relationship between $g_\chi$ and $f$,
\bea
	g_\chi^4 + (g_\chi^2 - f^2)^2 \cong \left\{
	\begin{array}{ll} 0.75,& {\rm MED,\,MAX}\\
		          0.64,& {\rm MIN}
	\end{array}\right. ~,
\label{gfrelic}
\eea
as shown in Fig.~\ref{fig:relic}.

From Fig.~\ref{fig:relic} we see that $g_\chi \cong 0.9$ for $f
\lesssim 0.8$, and therefore $g_\chi/m_{Z'}\cong 75$/TeV, in contrast
to the couplings of $Z'$ to the SM particles, $(g_q
g_l)^{1/2}/m_{Z'}\lesssim 1$/TeV.  This scenario thus requires a
substantial hierarchy $g_\chi \gtrsim 75\, (g_q, g_l)$, which might
require additional model-building to seem natural.  Here we defer such
questions and focus on the phenomenology.

\subsubsection{Sommerfeld enhancement}

At low temperatures $T< \delta m_\chi = m_{\chi_+}-m_{\chi_-}$, long
after freezeout, only the ground state DM $\chi_-$ is present: even
for very small mass splittings, the tree-level decay channel $\chi_+
\to \chi_- \nu_\mu\bar\nu_\mu$ by virtual $Z'$ emission is always
open.  The $\chi_-$ annihilation cross section at threshold is given
by Eq.~(\ref{chimxsect}).  For this to be large enough to give a
significant $\bar p$ signal, we need to be on the horizontal branch of
the relic density curves in Fig.\ \ref{fig:relic}, where $g_\chi \sim
0.75-0.9$.  This range corresponds to a cross section of
$(2.3-4.0)\times 10^{-26}$\,cm$^3/$s.

To match the central values needed for the AMS signal, we therefore
require respective Sommerfeld enhancement factors of order $S \sim
3,\,8,\,45$ for the MAX, MED, MIN propagation models.  To compute the
enhancement in the present model accurately could be complicated,
because it can generally be mediated both by $\phi$ and $Z'$ exchange,
and the latter interactions are inelastic.  

However it turns out that
this complication is avoided in our preferred region of parameter
space, because the DM mass splitting is so large that $Z'$ exchange is
suppressed.  Ref.\ \cite{Slatyer:2009vg} shows that the criterion for
neglecting Sommerfeld enhancement through $Z'$ exchange is $\delta
m_\chi > \alpha'^2 M_\chi/2 = (2.5-4)\,$GeV, where $\alpha' =
g_\chi^2/4\pi$.  Since $m_{Z'} = 2 g_\chi w$ and $\delta m_\chi = 2 f
w$, this puts a lower bound on the Yukawa coupling, $f \gtrsim
0.14-0.3$, which we will show is satisfied.  In contrast, dark Higgs
exchange proceeds through diagonal interactions with $\chi$, and since
$m_\varphi \ll m_\chi$, it can give rise to Sommerfeld-enhanced
annihilation despite the suppression of $Z'$ exchange.

\begin{figure}[t]
\hspace{-0.4cm}
\centerline{\includegraphics[width=0.9\hsize]{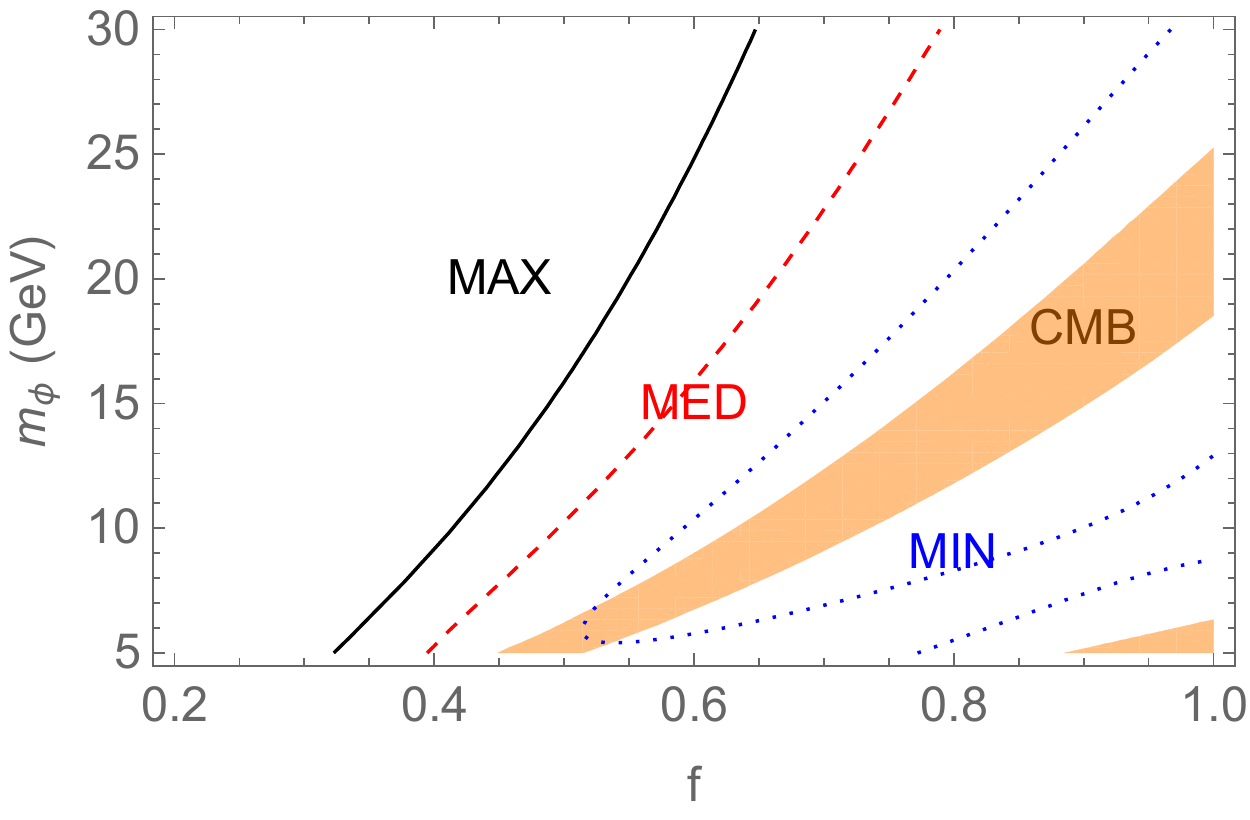}}
\caption{Values of $m_\varphi$ versus $f$ that give the observed
  antiproton excess at high energies, for the respective cosmic ray
  propagation models as labeled. The orange region is excluded by CMB constraints for DM 
with $m_\chi = 1800$\,GeV. For all curves $g_\chi$ is taken to be the value that gives the correct relic density (Eqn.\ \ref{gfrelic}). 
Where two values of $g_\chi$ give the correct relic density (see
fig.\ \ref{fig:relic} where $g_\chi$ can be double-valued), the larger one is used, since this requires a smaller
Sommerfeld enhancement for the galactic $\bar p$ signal.}
\label{enhance}
\end{figure}

We estimate the enhancement factor from $\phi$ exchange using
\cite{Cassel:2009wt}
\be
	S = \left|{\Gamma(a_+)\Gamma(a_-)/\Gamma(1+2iu)}\right|^2 ~,
\ee
where $a_\pm = 1+iu(1 \pm \sqrt{1 - x/u})$, $x=f^2/(16\pi \beta)$,
$\beta = v/c$, $u = 6\beta m_\chi/(\pi^2 m_\varphi)$, for dark matter
with velocity $v$ in the center-of-mass frame, which we take to be $v
= 10^{-3}c$.  The resulting correlated values of $m_\varphi$ and $f$
needed to fit the antiproton excess are shown in Fig.\ \ref{enhance}
for the three cosmic ray propagation models.  The required values of
$f$ are consistent with our assumption of sufficiently large DM mass
splittings (of order a few GeV) to justify the neglect of $Z'$
exchange in the enhancement factor, and $m_\varphi$ can be of the same
order as $m_{Z'}$ as expected.

Models with significant Sommerfeld enhancement are 
constrained by their potential to distort the cosmic microwave 
background (CMB) or disrupt big bang nucleosynthesis (BBN) 
\cite{Hisano:2011dc, Bringmann:2016din}. These effects can be significant
since the DM velocity is smaller during BBN and at 
recombination than at present in the Milky Way halo, possibly leading 
to a large enhancement of the annihilation cross section at those 
times. However, the Sommerfeld enhancement 
saturates at $v_{\rm min} \sim (m_\varphi/m_\chi) c$, which for the 
values of $m_\varphi$ and $m_\chi$ we consider above is 
$\sim 10^3$ km/s. 

In our scenario, DM kinetically decouples from the $Z'$ bosons when
they become nonrelativistic at a temperature of $T \sim m_{Z'}/3$.
The most probable velocity of the $\chi$ particles 
 is subsequently given by \cite{Cirelli:2016rnw}
\be
	v_0 \approx 10^{-8} \left( \frac{1 + z}{600} \right) \sqrt{\left(\frac{{\rm MeV}}{m_{Z'}}\right) \left(\frac{\rm GeV}{m_\chi} \right) } \, . 
\ee
For $m_\chi = 1800$ GeV and $m_{Z'} = 12$ GeV, $v_0 \sim 2 \times 10^{-12}$ m/s at $z = 600$, the redshift at which ionization due to DM annihilations can have the strongest effect on the CMB. As this is far below the saturation velocity, changes in DM velocity have little effect on the amount of Sommerfeld enhancement during this epoch, so we assume that $S$ is constant. 

With this approximation, we can use the 95\% CL limits on DM annihilation from the Planck collaboration \cite{Ade:2015xua}
\be
	S \langle \sigma v \rangle_{\chi \chi \to Z' Z'} f_{\rm eff} < 8.2 \times 10^{-28} \frac{\rm cm^3}{\rm s} \left( \frac{m_\chi}{\rm GeV} \right) \,. 
\ee
We take the efficiency parameter $f_{\rm eff}$ for annihilation to
$\bar b b$ from Ref.\ \cite{Slatyer:2015jla}. It has been shown in
Ref.\ \cite{Elor:2015bho} that limits from the CMB are insensitive to
whether one considers DM annihilating directly to $b$ quarks or to
mediators which cascade to $b$ quarks, as occurs in our model. The
limits from the CMB when $m_\chi = 1800$ GeV are shown in Fig.\
\ref{enhance}. In general the amount of Sommerfeld enhancement we need
to explain the $\bar p$ results is not enough to violate the CMB
bounds. Moreover current constraints from BBN are
weaker than those from the CMB, with observations of the ratio 
of deuterium to hydrogen constraining $\langle \sigma v \rangle 
\lesssim 1100 \times 10^{-26}$ cm$^3$/s at 95\% CL
\cite{Kawasaki:2015yya} for $m_\chi = 1800\,$GeV.

\subsubsection{Direct detection and collider constraints}

We avoid dark matter interactions with protons by $Z'$ exchange (due
to kinetic mixing) because of the highly inelastic nature of the
coupling $\bar\chi_+ \slashed{Z}'\chi_-$.  But the dark matter can
have a Higgs portal interaction from $\kappa |H|^2|\phi|^2$, allowing
the scalar $\phi$ to mix with the Higgs; the cross section on nucleons
is of order
\be
	\sigma_N \cong {(y_h\, f\, \theta\, m_N)^2\over \pi m_\varphi^4} ~,
\ee
where $y_h\cong 10^{-3}$ is the Higgs-nucleon coupling and $\theta\sim
\kappa v w/m_h^2$ is the mixing angle (with $v=246\,$GeV).  It can be
kept below current constraints by taking $f\theta\lesssim 10^{-3}$,
assuming that $m_\varphi\sim m_{Z'}$.  This implies $\kappa \lesssim
0.025$.

Our model escapes potentially stringent limits from monojets and
dijets \cite{Chala:2015ama} by its small couplings to quarks,
$g_q\lesssim 0.01$.  In the dimuon channel, limits on light $Z'$
bosons are significant if $g_q\sim g_l\sim 0.01$ for all flavors of
quarks \cite{Hoenig:2014dsa,Alves:2016cqf}, but these are relaxed for
our model which couples mainly to $b$ quarks.  A weak
constraint comes from the kinetic mixing coupling and its implications
for BaBar searches, electroweak precision data \cite{Hook:2010tw} and
proposed higher-energy collider searches.  The natural value of the
kinetic mixing parameter is of order $\epsilon \lesssim 5\times
10^{-4}$ (see
Eq.\ (\ref{kmix})), which is below the sensitivity of
BaBar
searches for $e^+e^-\to Z'\gamma$, $Z'\to e^+e^-,\mu^+\mu^-$
\cite{Lees:2014xha} (and our model is also slightly outside the mass 
range to which
they are sensitive,  $m_{Z'}<10.2\,$GeV).

Higher-mass regions can be probed in future
collider studies \cite{Curtin:2014cca}, but these also lack the 
sensitivity to probe such small $\epsilon$.  In contrast, the search for
Higgs decays $h\to Z' Z'\to 4\ell$ constrains the Higgs portal
coupling $\kappa |H|^2|\phi|^2$ to be $\kappa\lesssim 5\times 10^{-4}$
\cite{Castaneda-Miranda:2016qdk} , though this analysis only applies
for $m_{Z'} > 15\,$GeV, and would be slightly weakened by the branching ratio
for hadronic decay $Z'\to b\bar b$ in our model.  For such small
values of $\kappa$ the branching ratio for $h\to\varphi\varphi$ is of
order $(\kappa v/m_b)^2 \cong 10^{-3}$ and thus does not provide any
significant constraint.

\section{Conclusions}
\label{conclusions}

The observed anomalies in $B$-meson decays governed by $\bsll$ can be
explained if there is new physics in $\bsmumu$. In this paper we have
presented a model with a new $Z'$ vector boson that can explain the
anomalies. The model assumes that the SM flavor symmetries are gauged,
and that these symmetries are spontaneously broken, leaving only
$U(1)'$ at the TeV scale.  The $Z'$ is the gauge boson associated with
this $U(1)'$, and it couples only to left-handed third-generation
quarks and second-generation leptons in the flavor basis. When one
transforms to the mass basis, a $Z'$-mediated $\bsmumu$ decay is
generated. Taking into account all constraints on the model
($\bs$-$\bsbar$ mixing, $\bsnunubar$, neutrino trident production), we
show that the anomalous decays $B\to K\mu^+\mu^-$ can be explained.

Dark matter annihilation into $b$ quarks is a favored scenario for
indirect signals, making it natural to try to link it to anomalies in
$B$-meson decays.  We have demonstrated that, by
allowing the $Z'$ to also couple to \hbox{(quasi-)}Dirac dark matter $\chi$,
one can find a common explanation of the $\bsmumu$ anomalies and
tentative evidence for excess antiprotons in AMS-02 data. Two
alternative scenarios are interesting: a heavy $Z'$ and relatively
light $\chi$ to explain excess $\bar p$'s of energy $\sim 10\,$GeV,
and a light $Z'$ with heavy DM to generate $\bar p$'s at $\sim
300$\,GeV.  

Although we did not emphasize it, the heavy-$Z'$/light-DM scenario
has the
added advantage of also explaining the persistent gamma-ray excess
from the galactic center observed by Fermi-LAT
\cite{Daylan:2014rsa,Calore:2014xka,TheFermi-LAT:2015kwa}. 
Thanks to its suppressed couplings to light quarks, our model 
satisfies stringent limits from direct detection 
\cite{Hooper:2014fda,Escudero:2016kpw}.
Millisecond
pulsars have been suggested as an astrophysical origin for the gamma
ray excess, but it remains questionable whether they can plausibly account
for all of it \cite{Haggard:2017lyq}, leaving the dark matter
hypothesis as an interesting possibility.

Both of our proposed scenarios live in regions of parameter space that
make them imminently testable by a variety of experimental techniques.
The heavy-$Z'$/light-DM case requires couplings of $Z'$ that put it
close to bounds from $B_s$-$\bar B_s$ mixing, and to the sensitivity
of LHC searches for $Z'\to \mu^+\mu^-$.  In our model, lower than
usual $Z'$ masses
are allowed by LHC dilepton searches because of the invisible
branching ratio from $Z'$ decays to dark matter.
At the same time, the natural
one-loop level of kinetic mixing of $Z'$ with the photon implies that
the DM candidate is just below the current sensitivity
of direct detection searches.  For the light-$Z'$/heavy-DM case, a
light ($\sim 10\,$GeV) dark Higgs $\phi$ must also couple to the DM,
splitting the Dirac $\chi$ into Majorana particles with a large enough
mass splitting to be safe from direct detection. The coupling of
$\phi$ to the SM Higgs is already highly constrained by searches for
$h\to Z'Z'\to 4\mu$, suggesting that this is the most likely discovery
channel at colliders.

\bigskip
\noindent
{\bf Acknowledgments}: We thank Wolfgang Altmannshofer,
Kazunori Kohri, Matthew McCullough, Maxim Pospelov and Sam Witte for helpful
discussions or correspondence.  We also thank Grace Dupuis for
assistance with MadGraph.  This
work was financially supported by NSERC of Canada and by FQRNT of
Qu\'ebec (JMC, JMC).

\bibliographystyle{apsrev}
%\bibliography{ref_abelian_dm,scalarDM,seesaw,ref4thGen}

\end{document}